\documentclass[superscriptaddress,nofootinbib]{revtex4-1}
\usepackage{amsmath,amssymb}
\usepackage{graphicx}
\usepackage{braket}

\renewcommand{\d}{\ensuremath{\mathrm{d}}}

\renewcommand{\d}{\ensuremath{\mathrm{d}}}
\newcommand{\ii}{\ensuremath{\mathrm{i}}}
\newcommand{\p}{\partial}
\newcommand{\e}{\ensuremath{\mathrm{e}}}
\newcommand{\GZ}{\ensuremath{\mathrm{GZ}}}
\newcommand{\gf}{\ensuremath{\mathrm{gf}}}
\newcommand{\YM}{\ensuremath{\mathrm{YM}}}

\begin{document}

\title{{\bf Modeling the Gluon Propagator in Landau Gauge: \\[2mm]
            Lattice Estimates of Pole Masses and Dimension-Two Condensates
        }}

\author{A.~Cucchieri}
\email{attilio@ifsc.usp.br}
\affiliation{Ghent University, Department of Physics and Astronomy, Krijgslaan 281-S9, 9000 Gent, Belgium}
\affiliation{Instituto de F\'\i sica de S\~ao Carlos, Universidade de S\~ao Paulo, Caixa Postal 369, 13560-970 S\~ao Carlos, SP, Brazil}
\author{D.~Dudal}
\email{david.dudal@ugent.be}
\affiliation{Ghent University, Department of Physics and Astronomy, Krijgslaan 281-S9, 9000 Gent, Belgium}
\author{T.~Mendes}
\email{mendes@ifsc.usp.br}
\affiliation{Instituto de F\'\i sica de S\~ao Carlos, Universidade de S\~ao Paulo, Caixa Postal 369, 13560-970 S\~ao Carlos, SP, Brazil}
\author{N.~Vandersickel$\,$}
\email{nele.vandersickel@ugent.be}
\affiliation{Ghent University, Department of Physics and Astronomy, Krijgslaan 281-S9, 9000 Gent, Belgium}

\date{\today}
\begin{abstract}
We present an analytic description of numerical results for the
Landau-gauge SU(2) gluon propagator $D(p^2)$, obtained from lattice
simulations (in the scaling region) for the largest lattice sizes to
date, in d = 2, 3 and 4 space-time dimensions.
Fits to the gluon data in 3d and in 4d show very good agreement with
the tree-level prediction of the Refined Gribov-Zwanziger (RGZ) framework,
supporting a massive behavior for $D(p^2)$ in the infrared limit. In
particular, we investigate the propagator's pole structure and provide
estimates of the dynamical mass scales that can be associated with
dimension-two condensates in the theory. In the 2d case, fitting the data
requires a non-integer power of the momentum $p$ in the numerator of
the expression for $D(p^2)$.
In this case, an infinite-volume-limit extrapolation gives $D(0)=0$.
Our analysis suggests that this result is related to a particular symmetry
in the complex-pole structure of the propagator and {\em not} to purely
imaginary poles, as would be expected in the original Gribov-Zwanziger
scenario.
\end{abstract}
\maketitle

%%%%%%%%%%%%%%%%%%%%%%%%%%%%%%%%%%%%%%%%%%%%%%%%%%%%%%%%%%%%%%%%%%%%%%

\section{Introduction}

High-precision data from lattice simulations are a key ingredient
in our understanding of the low-energy aspects of Yang-Mills theories
associated with color confinement. In fact, whereas new insight into
the confinement mechanism may be gained by investigating the properties
of gauge-field configurations produced in the simulations (see e.g.\
\cite{Greensite:2003bk}), specific features of proposed confinement
scenarios may be tested by comparison with lattice data.
In this case, one may obtain physical values for a model's parameters
by fitting the predicted expression of a given observable to its
numerical realization.
One may also hope to over-constrain the proposed analytic forms, if the
fits can be done with sufficiently high number and wide range of data points,
from which systematic errors have been consistently eliminated.
In particular, this applies to predictions for the infrared behavior
of gluon and ghost propagators, formulated in Landau gauge for SU($N_c$)
gauge theory.
Here we perform a series of fits to the gluon propagator $D(p^2)$
and test the predictions of the so-called Refined Gribov-Zwanziger (RGZ)
framework, which differs from the scenario originally proposed by
Gribov \cite{Gribov:1977wm} and Zwanziger \cite{Zwanziger:1991gz}
through the introduction of dimension-two
condensates, associated with dynamical mass generation \cite{Dudal:2008sp}.
Our analysis is done for pure SU(2) gauge theory.
The data have been produced previously and discussed in
\cite{Cucchieri:2007md,Cucchieri:2007rg,Cucchieri:2008fc}
(see also \cite{Cucchieri:2010xr}), but they have not been
systematically fitted until now.
A companion paper with similar fits for the ghost propagator is under way
\cite{preparation}.
We note that an alternative comparison of these data to analytic
predictions was recently presented in \cite{Aguilar:2011yb}.

\vskip 3mm
The Gribov-Zwanziger confinement scenario is based on restricting the
functional integration to the first Gribov region $\Omega$
delimited by the first Gribov horizon $\partial \Omega$, where the smallest
nonzero (and positive) eigenvalue of the Faddeev-Popov matrix ${\cal M}$
goes to zero \cite{Gribov:1977wm,Zwanziger:1991gz}.
Let us recall that limiting the gauge configurations to this region was
an attempt --- made by Gribov in Ref.\ \cite{Gribov:1977wm} --- to fix
the gauge completely, getting rid of spurious gauge copies, known
thereafter as Gribov copies.
Now, $\Omega$ is a convex region of very high dimensionality and
therefore, as the infinite-volume limit is approached, the increase in
entropy should favor \cite{Zwanziger:1992qr} gauge configurations on the
surface $\partial \Omega$.
This in turn can cause the infrared enhancement of the ghost propagator
(which is related to the inverse of ${\cal M}$), inducing long-range effects
in the theory. Indeed, in Coulomb gauge, the restriction to the first
Gribov region causes the appearance of a confining color-Coulomb potential
\cite{Gribov:1977wm,Zwanziger:2002sh}.
Thus, in this scenario, formulated for momentum-space propagators, the
long-range features needed to explain the color-confinement mechanism are
manifest in the ghost propagator, whereas the momentum-space gluon propagator
$D(p^2)$ should be {\em suppressed} in the infrared limit. Such a suppression
is associated with violation of spectral positivity, which is commonly
interpreted as gluon confinement
\cite{Zwanziger:1991gz,Zwanziger:1989mf,Zwanziger:1993dh}.
In particular, $D(0)$ is originally
expected to be zero \cite{Gribov:1977wm,Zwanziger:1991gz}, corresponding to
maximal violation of spectral positivity.
The parametrization of this behavior as a
propagator having a pair of poles with purely imaginary masses has arisen
in the Gribov-Zwanziger approach \cite{Gribov:1977wm,Zwanziger:1991gz}, in
connection with the study of gauge copies.

Lattice studies (see \cite{Cucchieri:2010xr} for a recent review)
have confirmed the suppression of the gluon propagator in the infrared
limit and the enhancement of the ghost propagator at intermediate momenta.
However, considering lattice sizes large enough to allow
the investigation of the deep infrared regime, it is clear
that the results of the simulations
are {\em not} compatible with the scenario described above.
Indeed --- in space-time dimension d = 3, 4 --- the gluon propagator
shows a finite value as the momentum is taken to zero and the enhancement
of the ghost propagator is lost in this limit.
We note the very large lattice sizes employed in
order to observe such a behavior, $L \approx 20$ fm and larger
\cite{Cucchieri:2007md,Sternbeck:2007ug,Bogolubsky:2007ud,Cucchieri:2007rg,Cucchieri:2008fc,Bogolubsky:2009dc,Bornyakov:2009ug}.
In any case, violation of reflection positivity for the real-space gluon
propagator (see e.g.\ \cite{Cucchieri:2004mf}) is clearly observed in the data.

Recently, the quantitative description of the massive behavior for the
gluon propagator has been studied by several groups
\cite{Aguilar:2011yb,Fischer:2008uz,Binosi:2009qm,Iritani:2009mp,Bornyakov:2009ug,Dudal:2010tf,Aguilar:2010gm,Tissier:2010ts,Oliveira:2010xc,Aguilar:2010zx,Cucchieri:2009zt,Tissier:2011ey,Pennington:2011xs},
based on different proposed analytic forms.
Earlier attempts of fitting gluon-propagator data can be found for example
in Refs.\ \cite{Aiso:1997au,Leinweber:1998uu,Mandula:1999nj,Cucchieri:2003di}.
We note that some of these studies (see e.g.\
\cite{Aiso:1997au,Cucchieri:2003di}) have considered
the so-called Gribov-Stingl form \cite{Stingl:1985hx,Stingl:1994nk}
for modeling the massive behavior of the gluon propagator.
This form is a generalization of the Gribov propagator
described above, including pairs of complex-conjugate poles
with a nonzero real part.
As illustrated below, the behavior predicted for $D(p^2)$ in the RGZ
framework is also based on general complex-conjugate poles for the
(massive) propagator. This proposed form is given for SU($N_c$) gauge theory
and for four or three space-time dimensions
\cite{Dudal:2008sp,Dudal:2008rm,Vandersickel:2011ye,Vandersickel:2011zc,Dudal:2011gd}.
On the contrary, in the 2d case, the RGZ approach cannot be
implemented, since the dimension-two condensates would induce severe
infrared singularities, precluding the restriction of the functional
integration to the first Gribov region \cite{Dudal:2008xd}.
By fitting rational functions of $p^2$ to the {\em whole} range of
data for the SU(2) gluon propagator, we are able to obtain estimates for the
physical values of the masses in the RGZ framework, as well as to gain a
better understanding of the pole structure in the proposed expressions.
In each case, we look for the best fit to the data, with the smallest number
of independent parameters, and relate them to the condensates in the proposed
analytic forms only at the end.
Put differently, the predicted dependence of the fit parameters on the
condensates is {\em not} imposed in the fitting form, but is obtained as
a result of the fit.
This allows us to use a wide fitting range, considering all data points.
We note that predictions from the RGZ framework were already tested in
\citep{Dudal:2010tf}, showing good fits (using a somewhat different
analytic form and a smaller fitting range) to 4d lattice
data for the SU(3) case.

The paper is organized as follows. The Gribov-Zwanziger scenario
is briefly reviewed in Section \ref{sec:GZ}. The introduction of condensates
as part of the RGZ scenario is summarized is Section \ref{sec:RGZ},
where we present the expressions to be fitted to the lattice data.
The numerical results are discussed in general in
Section \ref{sec:runs} and in particular for the 4d, 3d and 2d cases
respectively in Sections \ref{sec:4d}, \ref{sec:3d} and \ref{sec:2d}.
We present our conclusions in Section \ref{conclusions}.

%%%%%%%%%%%%%%%%%%%%%%%%%%%%%%%%%%%%%%%%%%%%%%%%%%%%%%%%%%%%%%%%%%%%%%%%%%%

\section{The Gribov-Zwanziger Action}
\label{sec:GZ}

The Gribov-Zwanziger (GZ) action, introduced in 1989 \cite{Zwanziger:1989mf},
implements an all-order restriction of the path integral to the first
Gribov region
\begin{equation}\label{defgribovregion}
\Omega \; \equiv \; \left\{ \, A^a_{\mu}(x) : \,
\p_{\mu} A^a_{\mu}(x)=0 \, , \, \mathcal{M}^{ab}(x,y) > 0 \right\} \, ,
\end{equation}
where $A^a_{\mu}(x)$ is the gauge field and $\mathcal{M}^{ab}(x,y)$ is the
Landau-gauge Faddeev-Popov operator
\begin{equation}\label{mab2}
\mathcal{M}^{ab}(x,y) \; = \; - \delta(x-y) \, \p_\mu  D_\mu^{ab}
                      \; = \; \delta(x-y) \, \left( - \delta^{ab}\,\p_\mu^2
                   \,+\,  f^{abc} \, \p_\mu A_\mu^c \right) \, .
\end{equation}
By introducing auxiliary fields --- a pair of complex-conjugate
bosonic fields $\left( \overline{\varphi}_{\mu}^{ac},\,
\varphi_{\mu}^{ac} \right) $ and a pair of anticommuting complex-conjugate
fields $\left( \overline{\omega}_{\mu}^{ac},\, \omega_{\mu}^{ac}\right) $ ---
one is able to obtain a local renormalizable action \cite{Zwanziger:1992qr,Maggiore:1993wq,Dudal:2010fq}.
More precisely, the generating functional for the GZ action
can be written in d space-time dimensions as
\cite{Zwanziger:1989mf,Sobreiro:2004us,Dudal:2010fq}
\begin{equation}
Z(J) \; = \; \int [\d \Phi] \, \e^{S_\GZ + \int \d^{\rm d} x J_\mu^a (x)
A_\mu^a(x)} \,,
\label{eq:Z}
\end{equation}
where $S_\GZ$ is the local GZ action given by
\begin{equation}
\label{GZaction}
S_\GZ \; = \; S_0 \,+\, S_\gamma\,,
\end{equation}
with
\begin{equation}
S_{0} \; = \; S_{\mathrm{YM}} \,+\, S_{\gf} \,+\,
\int \d^{\rm d}x \, \left[\phantom{\frac{1}{2}}\!\!\!\!
\overline{\varphi}_{\mu}^{ac} \partial_{\nu} D_\nu^{ab}
\varphi_{\mu}^{bc} \,-\, \overline{\omega}_{\mu}^{ac} \partial_{\nu}
\left( D_\nu^{ab} \omega_{\mu}^{bc} \right)
\,-\, g \left( \partial_{\nu} \overline{\omega}_{\mu}^{an} \right) f^{abc} D_\nu^{bm} c^m \varphi_{\mu}^{cn} \right]
\end{equation}
and
\begin{equation}
S_{\gamma} \;=\; -\gamma^{2} g \int \d^{\rm d}x \left[ f^{abc} A_{\mu}^{a}
\varphi_{\mu}^{bc} \,+\, f^{abc} A_{\mu}^{a} \overline{\varphi}_{\mu}^{bc}
\,+\, \frac{{\rm d}}{g} \left( N_c^{2} - 1 \right) \gamma^{2} \right] \,.
\label{eq:Sofgamma}
\end{equation}
Here, $a,\,b,\,c,\,m$ and $n$ are color indices
in the adjoint representation,
$N_c$ is the number of colors, $\gamma$ is the so-called Gribov parameter,
$S_{\YM}$ is the classical Yang-Mills action
\begin{equation}
S_{\YM} \; = \; \frac{1}{4} \int \d^{\rm d}x \, F^a_{\mu\nu} F^a_{\mu\nu}
\end{equation}
and $S_\gf$ is the Landau-gauge-fixing action
\begin{equation}
S_{\gf} \; = \; \int \d^{\rm d}x \, \left( b^a \p_\mu A_\mu^a
                \,+\, \overline{c}^a \p_\mu D_\mu^{ab} c^b \right) \,,
\end{equation}
where the auxiliary field $b^a$ is a Lagrange multiplier enforcing
Landau gauge and (${\overline c}^a$, $c^a$) are the Faddeev-Popov ghost fields.
Also, we indicate with $[\d \Phi]$ the integration over all fields
$\Phi \in \left\{ A_\mu^a,\, c^a,\, \overline{c}^a,\, b^a,\,
\overline{\varphi}_{\mu}^{ac},\, \varphi_{\mu}^{ac},\, \omega_{\mu}^{ac},\,
\overline{\omega}_{\mu}^{ac} \right\}$.
Notice that one can simplify the notation of the auxiliary fields
$\left( \overline{\varphi}_\mu^{ac},\, \varphi_\mu^{ac},\,
\overline{\omega}_\mu^{ac},\, \omega_\mu^{ac}\right) $ in the action
$S_0$ using the symmetry of this action with respect
to the composite index $i \equiv \left( \mu,\, c \right)$. Thus, we can set
\begin{equation}
\left( \overline{\varphi}_\mu^{ac},\, \varphi_\mu^{ac},\,
\overline{\omega}_\mu^{ac},\, \omega_\mu^{ac} \right)
\; = \; \left( \overline{\varphi}_i^a,\, \varphi_i^a,\,
\overline{\omega}_i^a,\, \omega_i^a \right)
\end{equation}
and write
\begin{equation}
S_{0} \; = \; S_\YM \,+\, S_\gf \,+\, \int \d^{\rm d} x \,
\left[ \overline{\varphi}_i^a \,\p_\nu \left( D_\nu^{ab} \varphi^b_i \right)
\,-\, \overline{\omega}_i^a\, \p_\nu \left( D_\nu^{ab} \omega_i^b \right)
\,-\, g \left( \p_\nu \overline{\omega}_i^a \right) f^{abc} \,
D_\nu^{bm} c^m \varphi_i^c \right]  \, .
\end{equation}
Finally, the parameter $\gamma$ in Eq.\ (\ref{eq:Sofgamma}) is fixed by
the so-called gap equation (also known as the horizon condition)
\begin{equation}\label{horizonconditon2}
\braket{ \, g f^{abc} A^a_{\mu} \varphi^{bc}_{\mu} \, } \,+\,
\braket{ \, g f^{abc} A^a_{\mu} \overline{\varphi}^{bc}_{\mu} \, }
\,+\, 2 \, \gamma^2 {\rm d} \, (N_c^2 -1) \;=\; 0 \,,
\end{equation}
where $\braket{ \; }$ indicates the expectation value in the measure defined
by Eq.\ (\ref{eq:Z}). This condition is a consequence of the restriction of
the path integral to the first Gribov region.

As mentioned in the Introduction, one of the main outcomes of the GZ
theory is the modification of the behavior of gluon and ghost propagators in
the infrared (IR) limit in comparison with the perturbative behavior $1/p^2$
\cite{Gribov:1977wm,Zwanziger:1991gz,Zwanziger:1989mf,Zwanziger:1990by,Zwanziger:1992qr,Zwanziger:1993dh}.
Indeed, the gluon propagator becomes IR-suppressed with a tree-level behavior
given by
\begin{equation}\label{gluonprop2}
\Braket{ \, A^a_{\mu}(p) A^b_{\nu}(-p) \,}
\;\equiv\;
\delta^{ab} \, D(p^2) \,
\left( \, \delta_{\mu\nu} - \frac{p_{\mu} p_{\nu}}{p^2} \, \right)
\;=\; \delta^{ab} \,
\frac{p^2}{p^4 + 2 \, g^2 \, N_c \gamma^4}
\left( \, \delta_{\mu\nu} - \frac{p_{\mu} p_{\nu}}{p^2} \, \right)\,.
\end{equation}
This result is confirmed by one-loop calculations
\cite{Gracey:2007vv,Ford:2009ar}. The above expression for the
gluon propagator implies that $D(p^2)$ is null at zero momentum, which
in turn indicates maximal violation of
reflection positivity for the real-space gluon propagator $D(x)$
\cite{Zwanziger:1990by}. This violation is usually considered a
manifestation of gluon confinement
\cite{Zwanziger:1991gz,Zwanziger:1989mf,Zwanziger:1993dh}.
At the same time, one finds that the ghost propagator displays an
enhanced IR behavior
\begin{equation}
\Braket{ \, c^a(p) \overline{c}^b(p) \, }
\;\equiv\; \delta^{ab} \,{\mathcal G(p^2)}
\;\sim\; \delta^{ab} \, \frac{1}{p^4}\,.
\end{equation}
This behavior is indicative of a long-range interaction in the theory and it
should be related to quark confinement
\cite{Gribov:1977wm,Zwanziger:1993dh,Zwanziger:2010iz}.

%%%%%%%%%%%%%%%%%%%%%%%%%%%%%%%%%%%%%%%%%%%%%%%%%%%%%%%%%%%%%%%%%%%%%%%%

\section{The Refined Gribov-Zwanziger Framework}
\label{sec:RGZ}

More recently, the GZ action has been ``refined'' by taking into account
the possible existence of dimension-two condensates
\cite{Dudal:2008sp,Dudal:2008rm,Vandersickel:2011ye,Vandersickel:2011zc,Dudal:2011gd}. In the most general case
\citep{Dudal:2011gd}, four different condensates are considered, i.e.\
\begin{align}
\braket{A_\mu^a A_\mu^a}                                & \to -m^2  &
\Braket{\overline{\varphi}^a_i \varphi^a_{i}}           & \to M^2  &
\Braket{\varphi^a_i \varphi^a_{i}}                      & \to \rho &
\Braket{\overline{\varphi}^a_i \overline \varphi^a_{i}} & \to \rho^\dagger \; ,
\label{eq:condensates}
\end{align}
where we have listed the dynamical mass associated to each condensate.
(Note that $\rho$ is complex, whereas $\,-m^2$ and $M^2$ are real and positive.)
The condensate $\,-m^2$ is directly related to the gluon condensate
$\langle g^2 A^2 \rangle$ (see e.g.\ \cite{Dudal:2010tf}).
One can show \citep{Dudal:2011gd} that the Refined Gribov-Zwanziger (RGZ)
action can be renormalized. At the same time, there is clear evidence that
the original GZ theory dynamically transforms into the refined theory, since
the minimum of the associated effective potential favors non-vanishing
condensates \citep{Dudal:2011gd}. As displayed below, a non-zero value for these
condensates has an effect on the IR behavior of gluon propagators.
The effect on the ghost propagator will be discussed in a forthcoming
work \cite{preparation}.

%%%%%%%%%%%%%%%%%%%%%%%%%%%%%%%%%%%%%%%%%%%%%%%%%%%%%%%%%%%%%%%%%%%%%%%%%

\subsection{The Gluon Propagator}
\label{sec:RGZgluon}

In the presence of the four condensates considered in Eq.\
(\ref{eq:condensates}), the GZ gluon propagator (\ref{gluonprop2})
is modified \citep{Dudal:2011gd} as
\begin{equation}
D(p^2) \; = \;\frac{p^4 + 2 M^2 p^2+ M^4 - \rho \rho^\dagger}{
                    p^6 + p^4 \left(m^2 + 2 M^2 \right) + p^2 \left(2 m^2 M^2 + M^4 + \lambda^4 - \rho \rho^\dagger \right)
                  + m^2 \left( M^4 - \rho \rho^\dagger \right)
                  + M^2 \lambda^4 - \frac{\lambda^4}{2}
                  \left( \rho + \rho^\dagger \right)} \; ,
\label{eq:Drho}
\end{equation}
where the condensates $m^2$, $M^2$, $\rho$ are described above and
$\lambda^4$ is related to the Gribov parameter $\gamma$ through
$\lambda^4 = 2 g^2 N_c \gamma^4$. Since $\rho$ and $\rho^\dagger$ are
complex-conjugate quantities, we can set
\begin{eqnarray}
\rho &=& \rho_1 + \ii \rho_2 \nonumber\\
\rho^\dagger &=& \rho_1 - \ii \rho_2
\end{eqnarray}
and rewrite Eq.\ (\ref{eq:Drho}) as
\begin{equation}
D(p^2) \; = \; \frac{p^4 + 2 M^2 p^2 + M^4 - (\rho_1^2 + \rho_2^2)}{
          p^6 + p^4 \left (m^2 + 2 M^2 \right) + p^2 \left[2 m^2 M^2 + M^4
          + \lambda^4 - \left( \rho_1^2 + \rho_2^2 \right) \right]
          + m^2 \left[ M^4 - \left( \rho_1^2 + \rho_2^2
          \right) \right] + \lambda^4 \left( M^2 - \rho_1 \right)} \; .
\label{refinedgluonprop}
\end{equation}
It is interesting to notice that this propagator gets simplified if
$\rho = \rho^\dagger = \rho_1$ (i.e.\ $\rho_2=0$), which corresponds
to the equality $\Braket{ \, \overline{\varphi} \overline{\varphi} \, }
= \Braket{ \, \varphi \varphi \,}$ from (\ref{eq:condensates}).
Indeed, in this case one can factorize the quantity $p^2 + M^2 - \rho_1\,$
in the numerator and in the denominator of the above formula, obtaining
\begin{equation}\label{prop}
D(p^2) \; = \; \frac{p^2 + M^2 + \rho_1}{p^4 + p^2 \left( M^2 + m^2 +
      \rho_1 \right) + m^2 \left( M^2 + \rho_1 \right) +  \lambda^4} \;.
\end{equation}
Clearly, both propagators (\ref{refinedgluonprop}) and (\ref{prop}) have,
in principle, a finite {\em nonzero} value at zero momentum.
Nevertheless, if the value of $D(0)$ is sufficiently small, one still
finds that the real-space propagator $D(x)$ becomes negative for some (large)
value of $x$, i.e.\ reflection positivity can also be violated for these
propagators.

Note that both Eqs.\ \eqref{refinedgluonprop} and \eqref{prop} can be
decomposed as sums of propagators of the type $ \alpha / (p^2+\omega^2)$.
In particular, we can write Eq.\ \eqref{refinedgluonprop} as
\begin{equation}\label{gluonpropsimp}
D(p^2) \; = \; \frac{\alpha}{p^2+\omega_1^2} \,+\, \frac{\beta}{p^2+\omega_2^2}
\,+\, \frac{\gamma}{p^2+\omega_3^2} \;\,.
\end{equation}
To this end, we only need to solve the cubic equation
\begin{equation}
x^3 + x^2 \left (m^2 + 2 M^2 \right) + x \left[2 m^2 M^2 + M^4 + \lambda^4 -
\left( \rho_1^2 + \rho_2^2 \right) \right] +
m^2 \left[ M^4 - \left( \rho_1^2 + \rho_2^2 \right) \right] +
\lambda^4 \left( M^2 - \rho_1 \right) \; = \; 0 \; ,
\end{equation}
obtained by setting $p^2 = x\,$ in the denominator of
Eq.\ \eqref{refinedgluonprop}, and to find its three roots
$\omega_1^2$, $\omega_2 ^2$ and $\omega_3^2$. At the same time, the gluon
propagator in Eq.\ \eqref{prop} can be written as
\begin{equation}\label{4D3}
D(p^2) \; = \; \frac{\alpha_+}{p^2+\omega_{+}^2} \,+\,
\frac{\alpha_-}{p^2+\omega_{-}^2} \;,
\end{equation}
where we expect to have $\alpha_- = \alpha_+^*$ if
$\,\omega_{-}^2 = (\omega_{+}^2)^*$, i.e.\ if $\,\omega_{+}^2$ and
$\omega_{-}^2$ are complex conjugates.
Here, $\omega_{\pm}^2$ are the roots of the quadratic equation
\begin{equation}
x^2 + x \left( M^2 + m^2 + \rho_1 \right) + m^2 \left( M^2 + \rho_1 \right) +
\lambda^4 \; = \; 0 \; ,
\end{equation}
obtained by setting $p^2 = x$ in the denominator of Eq.\ \eqref{prop}.
Clearly, one finds complex-conjugate poles if
$\,|M^2 - m^2 + \rho_1| \,<\, 2\lambda^2$.

\vskip 3mm
Let us remark that rational forms such as \eqref{refinedgluonprop} and
\eqref{prop} for the gluon propagator were considered by Stingl
\cite{Stingl:1985hx,Stingl:1994nk}, as a way of accounting for
nonperturbative effects in an extended perturbative approach to Euclidean QCD.
More precisely, in his treatment, one expresses the proper vertices of
the theory as an iterative sequence of functions yielding a self-consistent
solution to the Dyson-Schwinger equations.
In particular, for the gluon propagator, this sequence is written
[see Eq.\ (2.10) in Ref.\ \cite{Stingl:1994nk}]
in terms of ratios of polynomials in the variable $p^2$, of degree $r$ in
the numerator and $r+1$ in the denominator, with $r = 0, 1, 2, \ldots\;$.
This functional form is then related, via operator product expansion, to
the possible existence of vacuum condensates of dimension $2n$, with $n\geq 1$.
At the same time, the associated complex-conjugate poles\footnote{See
\cite{Stingl:1994nk,Dokshitzer:2004ie,Alkofer:2003jj} for some considerations
concerning the issue of causality for propagators with complex poles.} are
interpreted as short-lived elementary excitations of the gluon field
\cite{Stingl:1985hx,Zwanziger:1991gz,Stingl:1994nk}.
By comparison, in the RGZ framework, one proposes specific forms for
the dimension-two condensates --- related to the auxiliary fields of
the GZ action --- and then obtains (at tree level) the rational
functions in Eqs.\ (\ref{refinedgluonprop}) and (\ref{prop}),
which correspond respectively to cases with $r = 3$ and 2 in Stingl's
iterative sequence.

\vskip 3mm
In Section \ref{sec:4d} below, we show that the simplest rational form
[with $r=2$, corresponding to Eq.\ (\ref{prop})]
works well in the 4d SU(2) case. A similar result was obtained for
the SU(3) case in \cite{Dudal:2010tf}.\footnote{Let us mention
that the condensate $\rho$ was not considered in Ref.\ \cite{Dudal:2010tf}.
Therefore, when discussing fit results using Eq.\ (\ref{prop}), we
must compare their values for $M^2$ to our values for $M^2 + \rho_1$.}
It may be noted that, in Ref.\ \cite{Aiso:1997au}, lattice data for the
4d SU(3) Landau-gauge gluon propagator were fitted using the above sequence
of functions with $r = 2, 4$ and it was found that a good description of
the data can be achieved only for $r=4$. Note, however, that the fit was
performed for the real-space propagator, for which the analysis is known
to be complicated by several technical issues
(see e.g.\ \cite{Cucchieri:2004mf,Bernard:1992hy}).
Moreover, although the lattice volume considered was rather large,
the study employed asymmetric lattices, which may give rise to
systematic effects \cite{Cucchieri:2006za}.

%%%%%%%%%%%%%%%%%%%%%%%%%%%%%%%%%%%%%%%%%%%%%%%%%%%%%%%%%%%%%%%%%%%%%%%%%%%

\section{Numerical Simulations}
\label{sec:runs}

The data presented here for the SU(2) Landau-gauge gluon propagator
were produced in 2007. The 3d and 4d cases were
run on the 4.5 Tflops IBM supercomputer at LCCA--USP \cite{lcca}, whereas
the 2d case was run on various PC clusters at the IFSC--USP.
Most of these data have already been discussed in Refs.\
\cite{Cucchieri:2007md,Cucchieri:2007rg,Cucchieri:2008fc,Cucchieri:2010xr},
but they were not systematically fitted up to now.

In the 4d case, we have considered lattice sides $N = $ 48, 56, 64, 80, 96
and 128, with lattice parameter $\beta = 2.2$.
The corresponding lattice spacing $a$ is approximately $0.210~\text{fermi}$,
implying that the smallest non-zero momentum
$p_{min} =  2 \sin(\pi / N)$ is about $46~\text{MeV}$ in physical units
for the $N = 128$ lattice. In this case, the physical lattice volume
$V = N^4$ is about $(27~\text{fermi})^4$. The number of gauge-field
configurations produced was 168 for $N = 128$ and about 250 for the other
lattice sizes.

In 3d, we have $N = $ 140, 200, 240 and 320 at $\beta = 3.0$, with
$a \approx 0.268~\text{fermi}$. Then, for the lattice volume $320^3$ the
smallest non-zero momentum $p_{min}$ is about $14~\text{MeV}$ and the
physical volume corresponds to about $(85~\text{fermi})^3$.  The number
of configurations was 630, 525, 350 and 125, respectively for the four
lattice sizes.

Finally, in the two-dimensional case, we considered $N = $ 80, 120, 160,
200, 240, 280 and 320 at $\beta = 10.0$.
In this case the lattice spacing is about $0.219~\text{fermi}$, the
lattice volumes $320^2$ correspond to $V \approx (70~\text{fermi})^2$
and in this case $p_{min} \approx 18~\text{MeV}$. We have about 600
configurations for each lattice volume.

In all cases we set the lattice spacing $a$ by considering the
input value $\sigma^{1/2} = 0.44~\text{GeV}$ for the string tension, which
is a typical value for this quantity in the 4d SU(3) case.
The evaluation of the lattice string tension is described in
\cite{Cucchieri:1995pn}, \cite{Cucchieri:2003di} and
\cite{Bloch:2003sk}, respectively for d = 2, 3 and 4. Note that all our
runs are in the scaling region \cite{Maas:2007uv,Cucchieri:2003di,Bloch:2003sk}
and all data refer to the SU(2) case.
Possible systematic effects due to Gribov copies
\cite{Cucchieri:1997dx,Silva:2004bv,
Bogolubsky:2005wf,Bogolubsky:2009qb,Maas:2009ph} as well as unquenching effects \cite{Furui:2005bu,Ilgenfritz:2006he,
Bowman:2007du,Boucaud:2001un} were {\em not} considered here.
Finite-volume effects, on the other hand, are well under control.
In particular, in 3d and in 4d, our largest lattice volumes
can be already considered as infinite. In the 2d case, a simple extrapolation
to infinite volume needs to be considered, as done in Section \ref{sec:2d}
to obtain the limiting value of $D(0)$.

Configurations have been generated by alternating heat-bath
updates of the link variables with micro-canonical steps, in order to reduce
the problem of critical slowing-down (see for example \cite{Cucchieri:2003zx}
and references therein). Gauge-fixing to Landau gauge was done using the
stochastic overrelaxation algorithm \cite{Cucchieri:1995pn,Cucchieri:2003zx}.
Let us also recall [see Eq.\ (\ref{gluonprop2})] that the gluon propagator
$D(p^2)$ in Landau gauge is evaluated using
\begin{equation}
D^{bc}_{\mu \nu}(p) \;=\; \sum_{x\mbox{,}\, y} \frac{e^{-2 \pi i \hat{p}
\cdot (x - y) / N}}{V}\, \langle A^b_{\mu}(x)\,A^c_{\nu}(y) \rangle \;=\;
\,\delta^{bc}\,\left(g_{\mu \nu}\,-\,\frac{p_{\mu}\,p_{\nu}}{p^2}\right)
D(p^2) \;.
\end{equation}
Here $A^b_{\mu}(x)$ is the lattice gluon field defined as\footnote{With
this definition of the lattice gluon field, the gluon propagator evaluated
on the lattice corresponds to the propagator $g^2 D(p^2)$ in the continuum,
which has mass dimension 2$-$d in the d-dimensional case
\cite{Cucchieri:1999sz}.}
\begin{equation}
A_{\mu}(x) \, = \, \frac{1}{2 i} \, \left[ \, U_{\mu}(x) \,-\, U_{\mu}^{\dagger}(x) \, \right] \; ,
\end{equation}
where $U_{\mu}(x)$ are the usual link variables of the Wilson action. Also, the momentum components $p_{\mu}$ are given by
\begin{equation}
p_{\mu} \;=\; 2 \, \sin\left(\frac{\pi \, \hat{p}_{\mu}}{N}\right)
\label{eq:kmu}
\end{equation}
and $\hat{p}_{\mu}$ takes values $0, 1, N-1$.

In 2d we considered momenta with components $(p,0)$ and $(p,p)$, plus all
possible permutations of the components. Similarly, in 3d, we have data for
momenta with components $(p,0,0)$, $(p,p,0)$ and $(p,p,p)$ and all possible
permutations of components. Finally, in 4d, we evaluated the propagator for
momenta with components $(p,0,0,0)$, $(p,p,0,0)$, $(p,p,p,0)$ and $(p,p,p,p)$.
In this case, we considered all possible permutations of the components for
momenta of the type $(p,0,0,0)$. On the contrary, we did not consider
permutations for the momenta $(p,p,p,0)$ and in the case $(p,p,0,0)$ we
allowed all permutations satisfying the constraint $p_4=0$. When permutations
of the momentum components were available, an average over the different
permutations was taken for each configuration.
In order to reduce discretization effects --- and in particular
those related to the breaking of rotational symmetry
\cite{Leinweber:1998uu,Ma:1999kn,deSoto:2007ht} --- we have considered, in
addition to the usual (unimproved) momentum defined by the squared
magnitude of the lattice momenta
\begin{equation}
p^2 \; = \; \sum_{\mu} \, p_{\mu}^2\,,
\label{eq:kunim}
\end{equation}
the improved definition \cite{Ma:1999kn}
\begin{equation}
p^2 \, = \, \sum_{\mu} \, p_{\mu}^2 \, + \, \frac{1}{12} \, \sum_{\mu}
\, p_{\mu}^4 \; .
\label{eq:kim}
\end{equation}
This definition does not affect the value of $p^2$ in the IR limit, but
modifies its value significantly for large momenta.
In particular, the largest value of $p^2$ --- obtained when
$\hat{p}_{\mu} = N/2$ in Eq.\ (\ref{eq:kmu}) for all
directions $\mu$ --- is given (in lattice units, for the d-dimensional case),
by 4d if the unimproved definition is considered, and by 16$\,$d/3 in
the improved case. For the $\beta$ values considered here, this implies that
the largest momentum $p_{max}$ is about $2.54, 2.55$ and $3.75~\text{GeV}$,
respectively in 2d, 3d and 4d in the unimproved case, and about $2.94, 2.94$
and $4.33~\text{GeV}$ using improved momenta.

In the next sections we present fits (obtained using {\tt gnuplot}) of the 4d,
3d and 2d data for the SU(2) gluon propagator and compare the fit results
to the predictions of the RGZ action, discussed above in Section
\ref{sec:RGZgluon}.
We remark that the shown data for $D(p^2)$ are {\em not} normalized.
Note that a renormalization condition at a given scale $\mu^2$ would
correspond to a rescaling of the overall factor $C$ in the fitting
forms considered below. The condensates and the poles, on the other hand,
are not affected by such a renormalization.
We also note that, whenever possible, we avoid rounding off the values of
the fit parameters. On the contrary, values for the associated physical
quantities (i.e.\ condensates and poles) are rounded to show errors with
one significant digit only. We refer to the {\tt gnuplot} documentation
\cite[``Statistical Overview'' section]{gnuplot}
for information on the significance of the standard errors calculated for
the fit parameters.

%%%%%%%%%%%%%%%%%%%%%%%%%%%%%%%%%%%%%%%%%%%%%%%%%%%%%%%%%%%%%%%%%%%%%%%%%%%%

\begin{figure}
\begin{center}
\includegraphics[width=.75\textwidth]{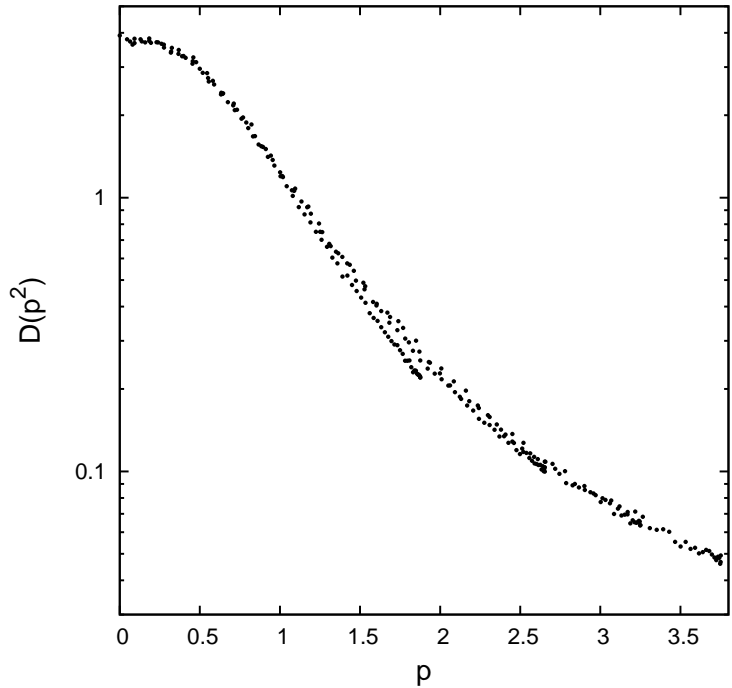}
\caption{Plot of the 4d gluon propagator $D(p^2)$ (in $\text{GeV}^{-2}$) as
a function of the (unimproved) momentum $p$ (in GeV) for the lattice volume
$V = 128^4$. As a consequence of the breaking of rotational invariance,
the data do not produce a smooth curve. Note the logarithmic scale on the
$y$ axis.}
\label{fig:4dgl-unimpr}
\end{center}
\end{figure}

\section{The 4d Case}
\label{sec:4d}

\begin{figure}
\begin{center}
\includegraphics[width=.75\textwidth]{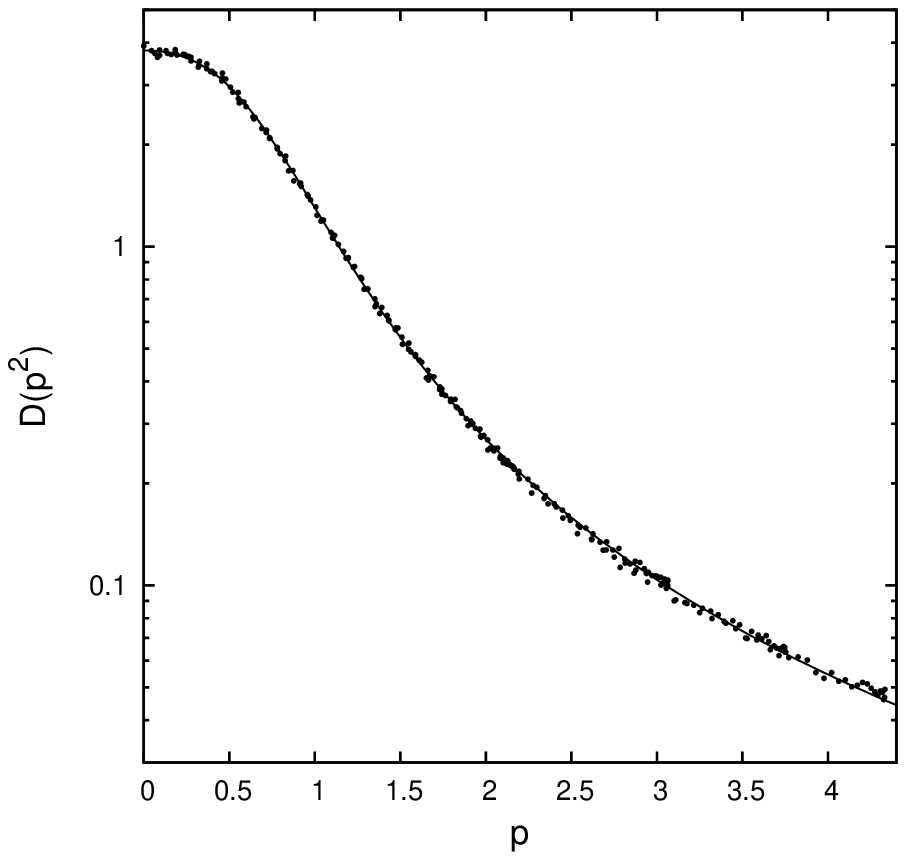}
\caption{Plot of the 4d gluon propagator $D(p^2)$ (in $\text{GeV}^{-2}$) as
a function of the (improved) momentum $p$ (in GeV) for the lattice volume
$V = 128^4$. We also show the fitting function $f_{1}(p^2)$
[see Eq.\ (\ref{eq:f4dgluon})] with the parameters reported in the last row of
Table \ref{tab:gluon4d-im}. Note the logarithmic scale on the $y$ axis.}
\label{fig:4dgl}
\end{center}
\end{figure}

As a first attempt in the 4d case, we consider a fitting
function of the simplest Gribov-Stingl form\footnote{Since our
largest momentum is of the order of $4 ~\text{GeV}$,
ultraviolet logarithmic corrections are not important to
describe the lattice data and they are not included in the fitting functions
proposed here. This also avoids the problem of having to regularize
the corresponding Landau pole by hand.}
\begin{equation}
f_{1}(p^2) \; = \; C \, \frac{p^2 + s}{p^4 + u^2 \, p^2 + t^2} \; ,
\label{eq:f4dgluon}
\end{equation}
which corresponds to the RGZ propagator in Eq.\ (\ref{prop}),
modulo a global rescaling factor $C$.
We note that, in order to improve the stability of the fit, we impose
some parameters to be positive, by setting them to be squares.
The results of the fit for all lattice volumes, using unimproved
and improved momenta, are reported respectively in Tables \ref{tab:gluon4d-unim}
and \ref{tab:gluon4d-im}. From the $\chi^2$/d.o.f.\ values one clearly
concludes that the use of improved momenta makes the behavior of the gluon
propagator smoother, allowing a better fit to the data.
This is also seen by comparing the data in Figs.\ \ref{fig:4dgl-unimpr}
and \ref{fig:4dgl}, plotted respectively for unimproved and improved momenta.
Let us stress that we are fitting the whole momentum range available and
that, for the largest lattice volume, we have 257 data points.

\begin{table}
\begin{center}
\begin{tabular}{cccccc}
\hline
\hline
  $ V $  &   $  C  $      &   $ u (\text{GeV})$   &  $t (\text{GeV}^2)$   & $ s (\text{GeV}^2)$   & $\chi^2$/d.o.f.\ \\
\hline
 $48^4$  &  0.567 (0.016) &  0.507 (0.065) &  0.607 (0.023) &  2.417 (0.269) & 16.47 \\
\hline
 $56^4$  &  0.572 (0.014) &  0.495 (0.058) &  0.602 (0.021) &  2.344 (0.234) & 15.46 \\
\hline
 $64^4$  &  0.566 (0.013) &  0.522 (0.052) &  0.612 (0.019) &  2.452 (0.223) & 15.00 \\
\hline
 $80^4$  &  0.562 (0.012) &  0.496 (0.048) &  0.612 (0.017) &  2.443 (0.199) & 16.59 \\
\hline
 $96^4$  &  0.567 (0.011) &  0.484 (0.044) &  0.604 (0.016) &  2.367 (0.178) & 15.96 \\
\hline
 $128^4$ &  0.560 (0.010) &  0.534 (0.037) &  0.621 (0.014) &  2.553 (0.166) & 10.65 \\
\hline
\hline
\end{tabular}
\end{center}
\caption{
Fits of the gluon-propagator data in the 4d case, for different lattice
volumes, using the fitting function $f_1(p^2)$ in Eq.\ (\ref{eq:f4dgluon})
and unimproved momenta [see Eq.\ (\ref{eq:kunim})].
We report, besides the value of the fit parameters, the $\chi^2$/d.o.f.\
obtained in each case. The whole range of momenta was considered for the fit.
Errors shown in parentheses correspond to one standard deviation.
}
\label{tab:gluon4d-unim}
\end{table}
\begin{table}
\begin{center}
\begin{tabular}{cccccc}
\hline
\hline
  $ V $  &   $  C  $      &   $ u (\text{GeV})$   &  $t (\text{GeV}^2)$   & $ s (\text{GeV}^2)$   & $\chi^2$/d.o.f.\ \\
\hline
 $48^4$  &  0.791 (0.007) &  0.755 (0.027) &  0.707 (0.013) &  2.419 (0.119) &  2.09 \\
\hline
 $56^4$  &  0.801 (0.006) &  0.734 (0.023) &  0.696 (0.012) &  2.305 (0.100) &  1.92 \\
\hline
 $64^4$  &  0.791 (0.007) &  0.760 (0.024) &  0.710 (0.012) &  2.425 (0.108) &  2.35 \\
\hline
 $80^4$  &  0.785 (0.005) &  0.734 (0.019) &  0.708 (0.009) &  2.404 (0.084) &  2.04 \\
\hline
 $96^4$  &  0.795 (0.004) &  0.717 (0.016) &  0.694 (0.008) &  2.291 (0.068) &  1.66 \\
\hline
 $128^4$ &  0.784 (0.005) &  0.768 (0.017) &  0.720 (0.009) &  2.508 (0.078) &  1.63 \\
\hline
\hline
\end{tabular}
\end{center}
\caption{
Fits of the gluon-propagator data in the 4d case, for different lattice
volumes, using the fitting function $f_1(p^2)$ in Eq.\ (\ref{eq:f4dgluon})
and improved momenta [see Eq.\ (\ref{eq:kim})].
We report, besides the value of the fit parameters, the $\chi^2$/d.o.f.\
obtained in each case. The whole range of momenta was considered for the fit.
Errors shown in parentheses correspond to one standard deviation.
}
\label{tab:gluon4d-im}
\end{table}

In order to extract the value of the condensates described in Section
\ref{sec:RGZgluon} above, we now consider only the fit results for the
volume $V=128^4$ (using improved momenta), reported in the last row of
Table \ref{tab:gluon4d-im}.
The corresponding plot is shown in Fig.\ \ref{fig:4dgl}.
By setting $f_{1}(p^2)$ equal to the RGZ propagator in Eq.\ (\ref{prop})
(modulo the global factor $C$) and using propagation of error,
we find for the condensates the values reported in the first column
of Table \ref{tab:gluon4d-param2}. (Note that, for this fitting form,
the condensates $M^2$ and $\rho_1$ cannot be determined separately.)
Let us mention that the values obtained here for $M^2+\rho_1$, $m^2$ and
$\lambda^4$ are in good quantitative agreement with the corresponding
values --- respectively indicated with $M^2$, $m^2$ and $ 2 g^2 N \gamma^4 $
--- reported in Ref.\ \cite{Dudal:2010tf} for the SU(3) case.\footnote{For
comparison with our values in Table \ref{tab:gluon4d-param2}, the
SU(3) condensates from \cite{Dudal:2010tf} are respectively
2.15(13) GeV$^2$, $-1.81(14)$ GeV$^2$ and 4.16(38) GeV$^4$.}
Also, as remarked above, the condensate $m^2$ may be used to obtain a value
for the gluon condensate $\langle g^2 A^2 \rangle$, through the relation
(see e.g.\ \cite{Dudal:2010tf})
\begin{equation}
\langle g^2 A^2 \rangle\;=\; - \frac{9}{13} \,\frac{N_c^2-1}{N_c} \, m^2\,.
\end{equation}
In our case, the value $m^2=-1.92(9)$ from Table \ref{tab:gluon4d-param2}
(using propagation of error) yields
$\langle g^2 A^2 \rangle = 1.99(9)$ GeV$^2$.

Furthermore, we verify from Table \ref{tab:gluon4d-param2}
that $\,|M^2-m^2+\rho_1| < 2\lambda^2$, justifying our
expectation (see end of Section \ref{sec:RGZgluon}) that the propagator
may be decomposed in terms of a pair of complex-conjugate poles.
We can thus write [see Eq.\ (\ref{4D3})]
\begin{equation}
f_{2}(p^2) \;=\; \frac{\alpha_+}{p^2+\omega_+^2} \,+\,
\frac{\alpha_-}{p^2 + \omega_-^2} \;=\;
\frac{2 a \,p^2 \,+\, 2(a v + b w)}{p^4 \,+\, 2 v \,p^2 \,+\, v^2 + w^2} \; ,
\label{eq:fpoles4d}
\end{equation}
with $\alpha_{\pm} = a \pm i b $ and $\omega_{\pm}^2 = v \pm i w $.
The results for the parameters $a, b, v$ and $w$ (again using propagation
of error) are shown in the first column of Table \ref{tab:gluon4d-param}.
Thus, the poles are complex conjugates whose imaginary part is more than
twice their real part. We recall that a Gribov propagator would have a null
real part.

In order to have better control of the errors on the values of the
condensates and poles, we have re-done the estimates described
above using a Monte Carlo error analysis (with 10000 samples).\footnote{To
this end, we considered independent Gaussian distributions
for the fit parameters. Thus, this Monte Carlo analysis may be considered
as a numerical check of the analytic propagation of error.} The corresponding
results are reported in the second columns of Table \ref{tab:gluon4d-param2}
and Table \ref{tab:gluon4d-param}.
Finally, we repeated the fit and the evaluation of the condensates and
poles using a bootstrap analysis (with 500 samples). In this case,
from the fit of $f_1(p^2)$ [see Eq.\ (\ref{eq:f4dgluon})] at $V = 128^4$,
we find the parameters
\begin{eqnarray}
C & = & 0.762 \pm 0.024 \,       \\
u & = & 0.755 \pm 0.035 \; \text{GeV}   \\
t & = & 0.698 \pm 0.027 \; \text{GeV}^2 \\
s & = & 2.292  \pm 0.253  \; \text{GeV}^2 \,.
\end{eqnarray}
The corresponding results for the condensates and poles of the propagator
are shown in the third columns of Tables \ref{tab:gluon4d-param2} and
\ref{tab:gluon4d-param}. Clearly, all results obtained agree within
one standard deviation.

\begin{table}
\begin{center}
\begin{tabular}{cccc}
\hline
\hline
parameter $\;$&$\;$ propagation of error $\;$ & $\;$ Monte Carlo analysis
$\;$ & $\;$ bootstrap analysis \\
\hline
$M^2+\rho_1\,(\text{GeV}^2)$     & $2.51(8)$  & $2.51(8)$  & $2.3(3)$  \\
\hline
$m^2\,(\text{GeV}^2)$       & $-1.92(9)$ & $-1.92(9)$ & $-1.7(2)$ \\
\hline
$\lambda^4\,(\text{GeV}^4)$ & $5.3(9)$   & $5.3(4)$   &  $4.5(9)$ \\
\hline
\hline
\end{tabular}
\end{center}
\caption{
Estimates of the parameters of the RGZ gluon propagator in Eq.\ (\ref{prop})
from fits (see last row of Table \ref{tab:gluon4d-im} above)
to the equivalent form $f_1(p^2)$ in Eq.\ (\ref{eq:f4dgluon}),
using propagation of error.
For comparison, we also report a Monte Carlo error analysis with 10000
samples and a bootstrap analysis (fit results given in the text) with
500 samples.
In all cases we considered the volume $V = 128^4$ and improved momenta.
Errors shown in parentheses correspond to one standard deviation.
}
\label{tab:gluon4d-param2}
\end{table}
\begin{table}
\begin{center}
\begin{tabular}{cccc}
\hline
\hline
parameter $\;$&$\;$ propagation of error $\;$ & $\;$ Monte Carlo analysis
$\;$ & $\;$ bootstrap analysis \\
\hline
$a$                   & 0.392(3)  &  0.392(2) &  0.38(1)  \\
\hline
$b$                   & 1.32(7)   &  1.32(5)  &  1.20(7)  \\
\hline
$v \,(\text{GeV}^2)$  & 0.29(2)   &  0.29(2)  &  0.29(3)  \\
\hline
$w \,(\text{GeV}^2)$  & 0.66(2)   &  0.66(1)  &  0.64(2)  \\
\hline
\hline
\end{tabular}
\end{center}
\caption{
Estimates of the parameters of the function $f_{2}(p^2)$ [see Eq.\
(\ref{eq:fpoles4d})]
from fits (see last row of Table \ref{tab:gluon4d-im} above)
to the equivalent form $f_1(p^2)$ in Eq.\ (\ref{eq:f4dgluon}),
using propagation of error.
For comparison, we also report a Monte Carlo error analysis with 10000
samples and a bootstrap analysis (fit results given in the text) with
500 samples.
In all cases we considered the volume $V = 128^4$ and improved momenta.
Errors shown in parentheses correspond to one standard deviation.
}
\label{tab:gluon4d-param}
\end{table}

As a second test, we have also tried to allow for the more general form
of the propagator, given in Eq.\ (\ref{refinedgluonprop}).
To this end, we consider the fitting function
\begin{equation}
f_{3}(p^2) \;=\; C \,
\frac{p^4 + 2 a^2 \, p^2 + b}{p^6 + c \, p^4 + d \, p^2 + e^2} \,,
\label{eq:f4dgluon-rgz}
\end{equation}
which has six parameters. As can be seen in Table \ref{tab:gluon4d-im-rgz},
the values of $\chi^2$/d.o.f.\ do not improve in comparison with the
previous (4-parameter) fit and, with the exception of the global factor $C$,
most of the parameters are determined with very large errors.
This suggests that the above function has too many (redundant) parameters,
making the fitting procedure quite unstable. Next, we reduce the number
of parameters by one and introduce a general form that will prove
useful in the description of the 3d data, in Section \ref{sec:3d}.
\begin{table}
\begin{center}
\begin{tabular}{cccccccc}
\hline
\hline
$V$ & $C$ & $d(\text{GeV}^4)$ & $e(\text{GeV}^3)$ & $b(\text{GeV}^4)$ &
$c(\text{GeV}^2)$  & $a(\text{GeV})$  &  $\chi^2$/d.o.f.\ \\
\hline
 $48^4$  &   0.889 (0.087) &  7.742 (9.663) &  2.469 (1.651) & 26.613 (32.940) & 9.692 (13.240) &  2.085 (1.143) &  1.95 \\
\hline
 $56^4$  &   0.798 (0.007) &  0.495 (0.020) &  0.055 (0.045) &  0.014 (0.023)  & 0.581 (0.057)  &  1.093 (0.030) &  2.04 \\
\hline
 $64^4$  &   0.795 (0.010) &  0.625 (0.662) &  0.336 (0.868) &  0.543 (2.790) &  0.733 (1.031)  &  1.121 (0.209) &  2.52 \\
\hline
 $80^4$  &   0.781 (0.006) &  0.514 (0.016) &  0.059 (0.037) &  0.016 (0.020) &  0.593 (0.046)  &  1.122 (0.025) &  2.08 \\
\hline
 $96^4$  &   0.893 (0.104) & 10.539 (14.750) & 3.036 (2.233) & 39.289 (53.150) & 16.558 (24.490) & 2.720 (1.693) &  1.56 \\
\hline
 $128^4$ &   0.784 (0.006) &  0.578 (2.192) &  0.229 (4.228) &  0.253 (9.335) &   0.691 (3.795) &  1.143 (0.843) &  1.69 \\
\hline
\hline
\end{tabular}
\end{center}
\caption{
Fits of the gluon-propagator data in the 4d case, for different lattice
volumes, using the fitting function $f_3(p^2)$ in Eq.\ (\ref{eq:f4dgluon-rgz})
and improved momenta [see Eq.\ (\ref{eq:kim})].
We report, besides the value of the fit parameters, the $\chi^2$/d.o.f.\
obtained in each case.
The whole range of momenta was considered for the fit.
Errors shown in parentheses correspond to one standard deviation.
}
\label{tab:gluon4d-im-rgz}
\end{table}
\begin{table}[bh]
\begin{center}
\begin{tabular}{ccccccc}
\hline
\hline
  $ V $  &   $  C  $   &   $ u (\text{GeV})$   &  $t (\text{GeV}^2)$   & $ s (\text{GeV}^2)$    & $k (\text{GeV}^2)$    & $\chi^2$/d.o.f.\ \\
\hline
$48^4$&   0.802 (0.009) &  0.686 (0.081) &  0.792 (0.030) &  1.662 (0.368) &  0.547 (0.149) &  2.02 \\
\hline
$56^4$&   0.809 (0.008) &  0.694 (0.063) &  0.761 (0.034) &  1.714 (0.365) &  0.622 (0.177) &  1.89 \\
\hline
$64^4$&   0.802 (0.008) &  0.701 (0.071) &  0.790 (0.028) &  1.716 (0.346) &  0.573 (0.144) &  2.28 \\
\hline
$80^4$&   0.793 (0.007) &  0.703 (0.046) &  0.767 (0.031) &  1.882 (0.308) &  0.668 (0.154) &  2.01 \\
\hline
$96^4$&   0.804 (0.006) &  0.673 (0.043) &  0.757 (0.022) &  1.694 (0.246) &  0.625 (0.119) &  1.62 \\
\hline
$128^4$ & 0.793 (0.006) &  0.727 (0.045) &  0.791 (0.025) &  1.903 (0.265) &  0.631 (0.120) &  1.60 \\
\hline
\hline
\end{tabular}
\end{center}
\caption{
Fits of the gluon-propagator data in the 4d case, for different lattice
volumes, using the fitting function $f_4(p^2)$ in Eq.\ (\ref{eq:fDvrgzsimple})
and improved momenta [see Eq.\ (\ref{eq:kim})].
We report, besides the value of the fit parameters, the $\chi^2$/d.o.f.\
obtained in each case.
The whole range of momenta was considered for the fit.
Errors shown in parentheses correspond to one standard deviation.
}
\label{tab:gluon4d-im-rgz2}
\end{table}
More precisely, we test the fitting function
\begin{equation}
f_{4}(p^2) \;=\; C \,\frac{\left(p^2 + s\right)\,\left(p^2 + 1\right)}{
\left(p^4 + u^2 \, p^2 + t^2\right)\,\left(p^2 + k\right)} \;=\; C \,
\frac{p^4 + (s+1) p^2 + s}{p^6 + (k+u^2) p^4 + (k u^2+t^2) p^2 + k t^2} \;.
\label{eq:fDvrgzsimple}
\end{equation}
This function is of the type (\ref{eq:f4dgluon-rgz}) (with different
parameters), but is written as a simple generalization of
$f_1(p^2)$ in Eq.\ (\ref{eq:f4dgluon}).\footnote{Note that a fit using
the more general form in Eq.\ (\ref{eq:f5}) below is unstable in
this case, yielding large errors for the fit parameters. Nevertheless,
this fit suggests the factor $(p^2+1)$ in the numerator of
(\ref{eq:fDvrgzsimple}), as adopted here.}
Fit results are shown in Table \ref{tab:gluon4d-im-rgz2}.
In this case the fits look reasonable.
The corresponding values for the condensates and $\lambda^4$ in
Eq.\ (\ref{refinedgluonprop})
are obtained by a Monte Carlo analysis (with 10000 samples) using the data
in the last row of Table \ref{tab:gluon4d-im-rgz2}.
We note that, in this case, the fitting form allows us to evaluate $M^2$,
$\rho_1$ and $|\rho|$ separately. We find the values
\begin{eqnarray}
M^2       &=&  1.5  \pm 0.1  \; \text{GeV}^2 \\
m^2       &=& -1.7  \pm 0.3  \; \text{GeV}^2 \\
\lambda^4 &=&  4.1  \pm 1.0  \; \text{GeV}^4 \\
\rho_1         &=&  0.5  \pm 0.1  \; \text{GeV}^2 \label{eq:parar} \\
\rho_1^2+\rho_2^2  &=&  0.2  \pm 0.1  \; \text{GeV}^2 \label{eq:parar2s2} \;.
\end{eqnarray}
We see that the errors are larger, and that the value of $\,M^2+\rho_1$
is incompatible with the numbers in Table \ref{tab:gluon4d-param2}
(obtained assuming $\rho_2=0$).
Also, a comparison of the values in (\ref{eq:parar}) and (\ref{eq:parar2s2})
suggests a very small (and imaginary) value for $\rho_2$, implying
that $\rho$ is real and thus supporting the simpler form in Eq.\ (\ref{prop}),
fitted above using the function $f_1(p^2)$.
Moreover, the $\chi^2$/d.o.f.\ is not better for the 5-parameter fit compared
to the 4-parameter fit, indicating that the latter is more stable.

We thus conclude that our best fit is $f_1(p^2)$ in Eq.\ (\ref{eq:f4dgluon}),
i.e.\ the 4d gluon-propagator lattice data favor the simplified expression
in Eq.\ (\ref{prop}), implying $\,\Braket{ \, \overline{\varphi}
\overline{\varphi} \, } = \Braket{ \, \varphi \varphi \,}$.

%%%%%%%%%%%%%%%%%%%%%%%%%%%%%%%%%%%%%%%%%%%%%%%%%%%%%%%%%%%%%%%%%%%

\section{The 3d Case}
\label{sec:3d}

\begin{figure}[t]
\begin{center}
\includegraphics[width=.75\textwidth]{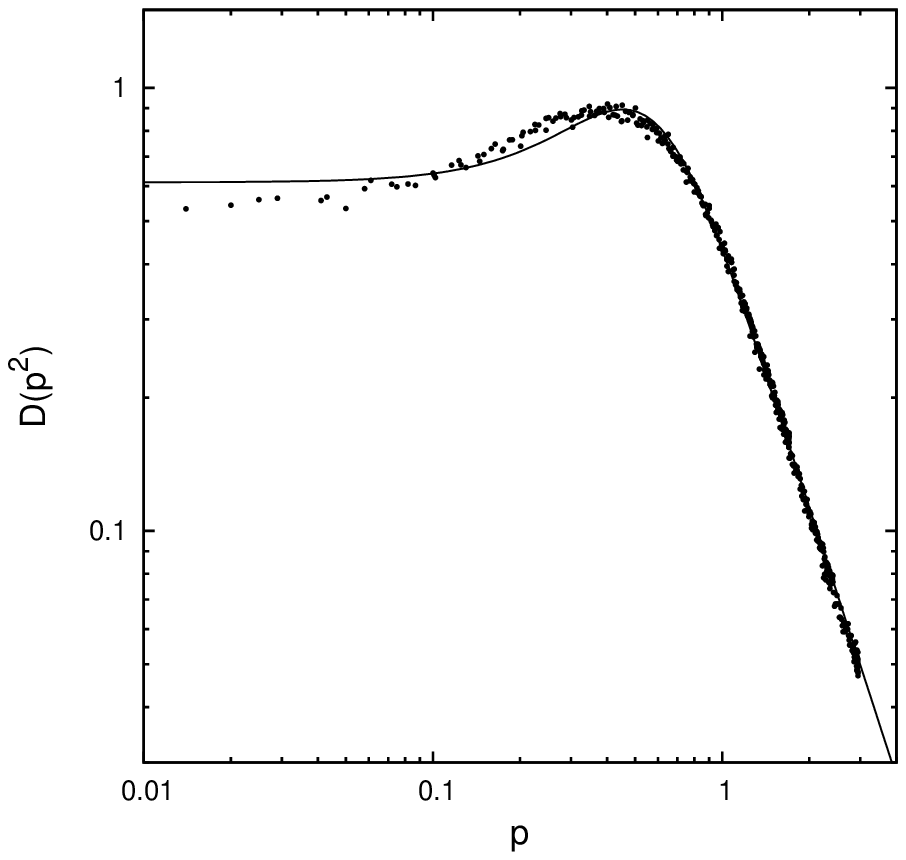}
\caption{Plot of the 3d gluon propagator $D(p^2)$ (in $\text{GeV}^{-1}$) as
a function of the (improved) momentum $p$ (in GeV) for the lattice volume
$V = 320^3$. We also show the fitting function $f_{1}(p^2)$ [see
Eq.\ (\ref{eq:f4dgluon})] with the parameters reported in the last row of
Table \ref{tab:gluon3d-im-not}. Note the logarithmic scale on both axes.}
\label{fig:3dgl-rgz}
\end{center}
\end{figure}

\begin{figure}
\begin{center}
\includegraphics[width=.75\textwidth]{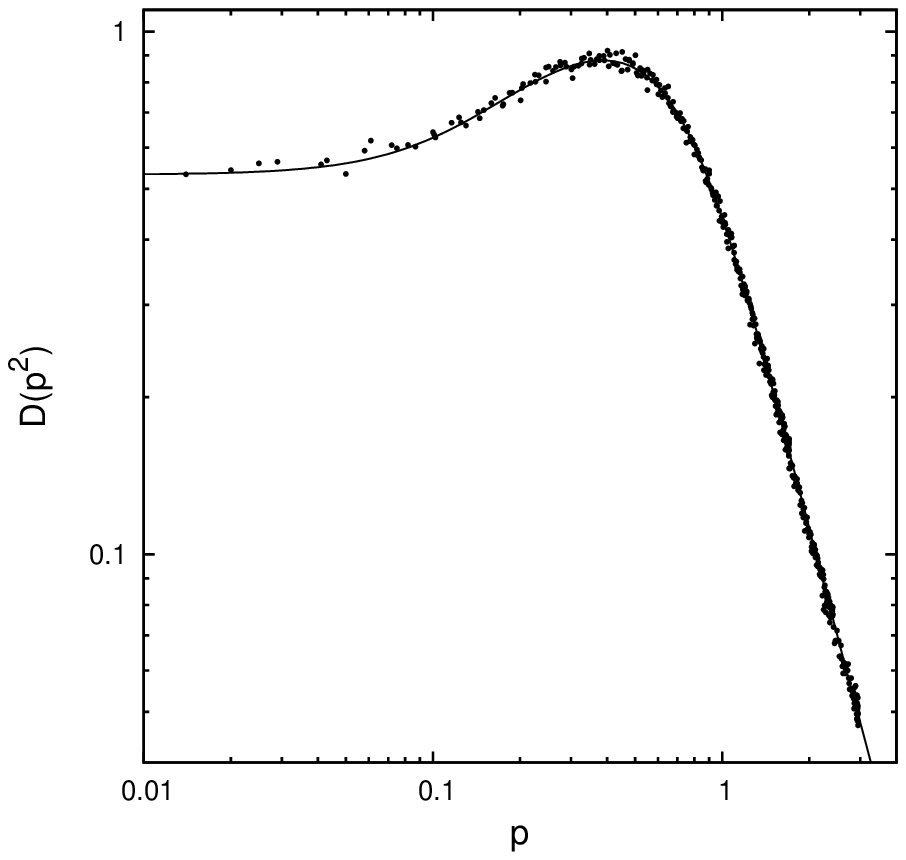}
\caption{Plot of the 3d gluon propagator $D(p^2)$ (in $\text{GeV}^{-1}$) as
a function of the (improved) momentum $p$ (in GeV) for the lattice volume
$V = 320^3$. We also show the fitting function $f_{4}(p^2)$ [see
Eq.\ (\ref{eq:fDvrgzsimple})] with the parameters reported in the last row of
Table \ref{tab:gluon3d-im}. Note the logarithmic scale on both axes.}
\label{fig:3dgl}
\end{center}
\end{figure}

In this case the simplified fitting form $f_{1}(p^2)$ in
Eq.\ (\ref{eq:f4dgluon}) is not able to describe well the lattice data.
Indeed, even using improved momenta (see Table \ref{tab:gluon3d-im-not}),
the $\chi^2$/d.o.f.\ values obtained are quite large. Moreover, as can
be seen in Fig.\ \ref{fig:3dgl-rgz}, the fit clearly fails in the IR
region.\footnote{In order to highlight the results at small momenta, here
and in Fig.\ \ref{fig:3dgl} we present the plot with a logarithmic scale
on both axes.}
The situation improves by considering the (5-parameter) fitting function
$f_4(p^2)$ in Eq.\ (\ref{eq:fDvrgzsimple}) above, as can be seen from the
results reported in Tables \ref{tab:gluon3d-unim} and \ref{tab:gluon3d-im},
obtained respectively using unimproved and improved momenta.
Note that, as in the 4d case, the use of improved momenta helps to
obtain a better fit to the data.

\begin{table}
\begin{center}
\begin{tabular}{cccccc}
\hline
\hline
  $ V $  &  $C (\text{GeV})$     &   $ u (\text{GeV})$   &  $t (\text{GeV}^2)$   & $ s (\text{GeV}^2)$   & $\chi^2$/d.o.f.\ \\
\hline
$140^3$  &  0.441 (0.003) &  0.306 (0.013) &  0.385 (0.009) &  0.217 (0.013) & 10.86 \\
\hline
$200^3$  &  0.440 (0.002) &  0.305 (0.011) &  0.389 (0.008) &  0.223 (0.011) &  8.68 \\
\hline
$240^3$  &  0.443 (0.002) &  0.307 (0.010) &  0.374 (0.007) &  0.198 (0.009) &  6.53 \\
\hline
$320^3$  &  0.445 (0.002) &  0.296 (0.011) &  0.365 (0.006) &  0.183 (0.008) &  3.19 \\
\hline
\hline
\end{tabular}
\end{center}
\caption{
Fits of the gluon-propagator data in the 3d case, for different lattice
volumes, using the fitting function $f_1(p^2)$ in Eq.\ (\ref{eq:f4dgluon})
and improved momenta [see Eq.\ (\ref{eq:kim})].
We report, besides the value of the fit parameters, the $\chi^2$/d.o.f.\
obtained in each case.
The whole range of momenta was considered for the fit.
Errors shown in parentheses correspond to one standard deviation.
}
\label{tab:gluon3d-im-not}
\end{table}

\begin{table}
\begin{center}
\begin{tabular}{ccccccc}
\hline
\hline
  $ V $ &  $C (\text{GeV})$    &   $ u (\text{GeV})$   &  $t (\text{GeV}^2)$   &
                                       $ s (\text{GeV}^2)$   & $ k (\text{GeV}^2)$   & $\chi^2$/d.o.f.\ \\
\hline
$140^3$ & 0.289 (0.002) &  0.382 (0.022) &  0.552 (0.006) &  0.018 (0.003) &  0.030 (0.006) & 10.48 \\
\hline
$200^3$ & 0.289 (0.002) &  0.386 (0.019) &  0.552 (0.006) &  0.019 (0.003) &  0.032 (0.006) &  9.45 \\
\hline
$240^3$ & 0.290 (0.002) &  0.393 (0.017) &  0.550 (0.005) &  0.020 (0.003) &  0.034 (0.005) &  6.55 \\
\hline
$320^3$ & 0.290 (0.002) &  0.389 (0.017) &  0.549 (0.005) &  0.019 (0.003) &  0.035 (0.005) &  2.89 \\
\hline
\hline
\end{tabular}
\end{center}
\caption{
Fits of the gluon-propagator data in the 3d case, for different lattice
volumes, using the fitting function $f_4(p^2)$ in Eq.\ (\ref{eq:fDvrgzsimple})
and unimproved momenta [see Eq.\ (\ref{eq:kunim})].
We report, besides the value of the fit parameters, the $\chi^2$/d.o.f.\
obtained in each case.
The whole range of momenta was considered for the fit.
Errors shown in parentheses correspond to one standard deviation.
}
\label{tab:gluon3d-unim}
\end{table}
\begin{table}
\begin{center}
\begin{tabular}{ccccccc}
\hline
\hline
  $ V $ &  $C (\text{GeV})$    &   $ u (\text{GeV})$   &  $t (\text{GeV}^2)$   &
                                       $ s (\text{GeV}^2)$   & $ k (\text{GeV}^2)$   & $\chi^2$/d.o.f.\ \\
\hline
$140^3$ &  0.407 (0.001) &  0.654 (0.008) &  0.623 (0.004) &  0.022 (0.002) &  0.041 (0.003) &  2.14 \\
\hline
$200^3$ &  0.407 (0.001) &  0.655 (0.007) &  0.623 (0.004) &  0.024 (0.002) &  0.043 (0.003) &  1.92 \\
\hline
$240^3$ &  0.408 (0.001) &  0.662 (0.007) &  0.620 (0.004) &  0.025 (0.002) &  0.047 (0.003) &  1.59 \\
\hline
$320^3$ &  0.408 (0.001) &  0.656 (0.008) &  0.619 (0.005) &  0.023 (0.002) &  0.046 (0.004) &  1.19 \\
\hline
\hline
\end{tabular}
\end{center}
\caption{
Fits of the gluon-propagator data in the 3d case, for different lattice
volumes, using the fitting function $f_4(p^2)$ in Eq.\ (\ref{eq:fDvrgzsimple})
and improved momenta [see Eq.\ (\ref{eq:kim})].
We report, besides the value of the fit parameters, the $\chi^2$/d.o.f.\
obtained in each case.
The whole range of momenta was considered for the fit.
Errors shown in parentheses correspond to one standard deviation.
}
\label{tab:gluon3d-im}
\end{table}

\begin{table}
\begin{center}
\begin{tabular}{cccc}
\hline
\hline
 parameter  $\;$&$\;$ propagation of error $\;$ & $\;$ Monte Carlo analysis $\;$ & $\;$ bootstrap analysis \\
\hline
 $M^2 \,(\text{GeV}^2)$  & 0.512 (1) & 0.512 (1) & 0.513 (1) \\
\hline
 $m^2 \,(\text{GeV}^2)$  & $-0.55 (1)$ & $-0.55 (1)$ & $-0.52 (2)$ \\
\hline
 $\lambda^4 \,(\text{GeV}^4)$ & 0.94 (1) & 0.94 (1) & 0.91 (3) \\
\hline
 $\rho_1 \,(\text{GeV}^2)$ & 0.479 (2) & 0.479 (2) & 0.477 (2) \\
\hline
 $\rho_2 \,(\text{GeV}^2)$ & 0.09 (1) & 0.094 (9) & 0.100 (6)    \\
\hline
\hline
\end{tabular}
\end{center}
\caption{
Estimates of the parameters of the RGZ propagator in
Eq.\ (\ref{refinedgluonprop}) from fits (see last row of
Table \ref{tab:gluon3d-im} above) to the equivalent form $f_4(p^2)$
in Eq.\ (\ref {eq:fDvrgzsimple}).
Errors are obtained using propagation of error, a Monte Carlo analysis with
10000 samples and a bootstrap analysis with 500 samples. In all cases we
considered the volume $V = 320^3$ and improved momenta.
Errors shown in parentheses correspond to one standard deviation.
}
\label{tab:gluon3d-param2}
\end{table}
\begin{table}[b]
\begin{center}
\begin{tabular}{cccc}
\hline
\hline
 parameter  $\;$&$\;$ propagation of error $\;$ & $\;$ Monte Carlo analysis $\;$ & $\;$ bootstrap analysis \\
\hline
 $\alpha \,(\text{GeV})$ & $-0.024 (5)$ & $-0.024 (5)$ & $-0.029 (4)$  \\
\hline
 $\omega_1^2 \,(\text{GeV}^2)$ & 0.046 (4) & 0.046 (4) & 0.046 (4) \\
\hline
 $a \,(\text{GeV})$ & 0.216 (3) & 0.216 (2) & 0.220 (4)  \\
\hline
 $b \,(\text{GeV})$ & 0.27 (5) & 0.271 (3) & 0.275 (3) \\
\hline
 $v \,(\text{GeV}^2)$ & 0.215 (5) & 0.215 (5) & 0.23 (1) \\
\hline
 $w \,(\text{GeV}^2)$ & 0.580 (6) & 0.580 (6) & 0.57 (1) \\
\hline
\hline
\end{tabular}
\end{center}
\caption{
Estimates of the parameters of the function $f_{6}(p^2)$ [see
Eq.\ (\ref{eq:fpoles3d})] from fits (see last row of
Table \ref{tab:gluon3d-im} above) to the equivalent form $f_4(p^2)$
in Eq.\ (\ref {eq:fDvrgzsimple}).
Errors are obtained using propagation of error, a Monte Carlo analysis with
10000 samples and a bootstrap analysis with 500 samples. In all cases we
considered the volume $V = 320^3$ and improved momenta.
Errors shown in parentheses correspond to one standard deviation.
}
\label{tab:gluon3d-param}
\end{table}

One can also try to use the more general function
\begin{equation}
f_{5}(p^2) \;=\; C\, \frac{\left(p^2 + s\right) \, \left(p^2 + l\right)}{
            \left(p^4 + u^2 \, p^2 + t^2\right) \, \left(p^2 + k\right)} \,,
\label{eq:f5}
\end{equation}
obtained by introducing the extra parameter $l$. In this case, the results
of the fit using improved momenta for the lattice volume $V=320^3$ are
\begin{eqnarray}
C & = & 0.405 \pm 0.003  \; \text{GeV} \\
u & = & 0.692 \pm 0.040  \; \text{GeV} \\
t & = & 0.635 \pm 0.018  \; \text{GeV}^2 \\
s & = & 0.025 \pm 0.002  \; \text{GeV}^2 \\
k & = & 0.050 \pm 0.007  \; \text{GeV}^2 \\
l & = & 1.092 \pm 0.103  \; \text{GeV}^2
\end{eqnarray}
with $\chi^2$/d.o.f.\ $=$ 1.19. By noticing that $l \approx 1$ and by
comparing these values to the corresponding ones from the fit using
$f_4(p^2)$ in Eq.\ (\ref{eq:fDvrgzsimple}) (reported in the last row of
Table \ref{tab:gluon3d-im}), it is clear that the two results are
equivalent. At any rate, the values of $\chi^2$/d.o.f.\ for
the 6-parameter fit [using the function $f_5(p^2)$ in Eq.\ (\ref{eq:f5})]
and for the fit in Table \ref{tab:gluon3d-im} are the same,
indicating that the latter is more stable.

In order to evaluate the condensates of the RGZ model, we thus consider only
the results from the fit using $f_4(p^2)$, given for the lattice size $N=320$
in the last row of Table \ref{tab:gluon3d-im}. (The corresponding plot is
shown in Fig.\ \ref{fig:3dgl}.)
By setting $f_{4}(p^2)$ [see Eq.\ \ref{eq:fDvrgzsimple})] equal to the RGZ
propagator (\ref{refinedgluonprop}) modulo the global factor $C$, we find
(using propagation of error) the values for the condensates reported in the
first column of Table \ref{tab:gluon3d-param2}.
Note that, using this fitting form, we are able to evaluate $M^2$, $\rho_1$
and $|\rho|$ (and therefore $\rho_2$) separately.
In this case, we can see that $\rho_2 \neq 0$ and $\rho$ is indeed a complex
quantity. This is consistent with the fact that the (four-parameter) fit to
the simplified form $f_1(p^2)$ fails, as seen above.

Finally, we decompose the propagator as in Eq.\ (\ref{gluonpropsimp}) with
$\beta = a + i b $, $\gamma = a - i b $, $\omega_2^2 = v + i w $ e
$\omega_3^2 = v - i w $, i.e.\ we consider the function
\begin{equation}
f_{6}(p^2) \; = \;
\frac{\alpha}{p^2 + \omega_1^2} \, + \,
\frac{2 a \, p^2 \,+\, 2(a v + b w)}{p^4 \,+\, 2 v\, p^2 \,+\, v^2 + w^2} \,.
\label{eq:fpoles3d}
\end{equation}
We find (again using propagation of error) the results\footnote{Clearly, we
have $\omega_1^2 = k$ from $f_4(p^2)$.} reported in the first column of Table
\ref{tab:gluon3d-param}. Note that the imaginary part $w$ of the
complex-conjugate poles is more than twice the value of their real part $v$,
as in the 4d case.
Note also that the mass $\omega_1$ and the residue $\alpha$
associated with the real pole are very small. 
Moreover, $\alpha$ is negative, which may
be associated with violation of reflection positivity, indicating that
this mass cannot correspond to a physical degree of freedom.

Also in this case we have repeated the analysis using a Monte Carlo estimate
for the errors (with 10000 samples) and the bootstrap method (with 500 samples).
The results are shown, respectively, in the second and in the third columns of
Tables \ref{tab:gluon3d-param2} and \ref{tab:gluon3d-param}. The values of the
fit parameters for the function (\ref{eq:fDvrgzsimple}) using the bootstrap
method for the lattice volume $V = 320^3$ are
\begin{eqnarray}
C & = & 0.411 \pm 0.004 \; \text{GeV}   \\
u & = & 0.673  \pm 0.016  \; \text{GeV}   \\
t & = & 0.611 \pm 0.006 \; \text{GeV}^2 \\
s & = & 0.025 \pm 0.002 \; \text{GeV}^2 \\
k & = & 0.052 \pm 0.005 \; \text{GeV}^2 \; ,
\end{eqnarray}
which should be compared to the results shown in the last row of Table
\ref{tab:gluon3d-im}. Again, all results obtained using the three different
analyses agree within one standard deviation.

\vskip 3mm
We thus conclude that, in the 3d case, the data support a gluon propagator
given by the general RGZ form (\ref{refinedgluonprop}), in which the
condensate $\rho$ is a complex quantity. This is in contrast with the 4d
case seen in the previous section, for which $\rho$ was real. There are
also significant differences for the values of the other condensates
and of $\lambda^4$ in comparison with the 4d case.
The masses from the complex-conjugate poles, on the contrary, have similar
values in 3d and 4d (see Tables \ref{tab:gluon3d-param} and
\ref{tab:gluon4d-param} respectively).

%%%%%%%%%%%%%%%%%%%%%%%%%%%%%%%%%%%%%%%%%%%%%%%%%%%%%%%%%%%%%%%%%%%%%%%%%%

\section{The 2d Case}
\label{sec:2d}

\begin{figure}
\begin{center}
\includegraphics[width=.75\textwidth]{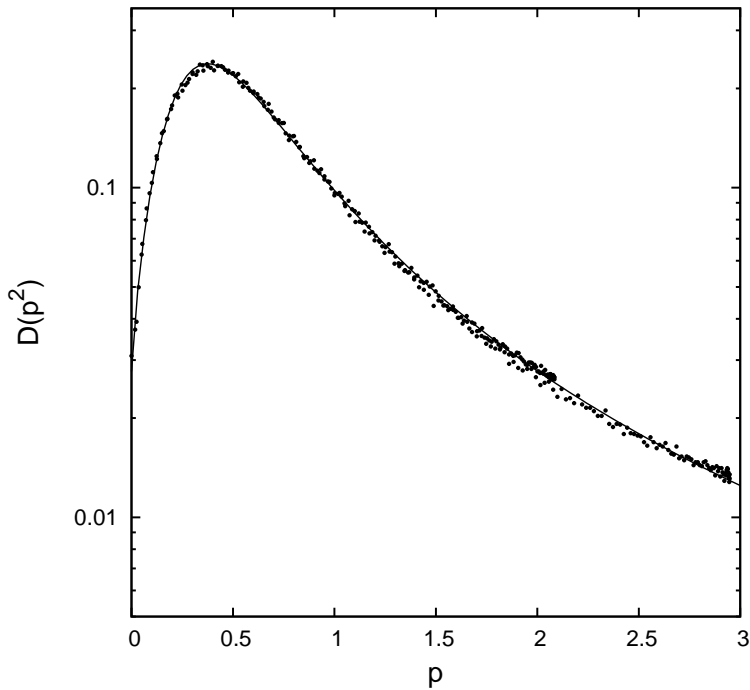}
\caption{Plot of the 2d gluon propagator $D(p^2)$ as a function of the
(improved) momentum $p$ (in GeV) for the lattice volume $V = 320^2$. We also
show the fitting function $f_{7}(p^2)$ [see Eq.\ (\ref{eq:f2dgluon})] with
the parameters reported in the last row of Table \ref{tab:gluon2d-im}.
Note that in 2d, with our convention, the gluon propagator $D(p^2)$ is
dimensionless. Also note the logarithmic scale on the $y$ axis.}
\label{fig:D2d}
\end{center}
\end{figure}

In the two-dimensional case the situation is different. Indeed, we know
from Refs.\ \cite{Cucchieri:2007rg,Maas:2007uv,Cucchieri:2011um} that the
gluon propagator at zero momentum $D(0)$ does go to zero in the
infinite-volume limit, even though one always has $D(0) > 0$ at finite
lattice volume.  Moreover,
the behavior of $D(p^2)$ at small momenta is of the type $p^{\eta}$, with some
non-integer power ${\eta} \approx 0.8$. This makes the fitting procedure more
complicated than in the above cases, for which polynomial forms were used.
After trying several generalizations of the fitting functions considered
in the 4d and 3d cases, we found that a good fit to the gluon data can be
obtained using the function
\begin{equation}
f_{7}(p^2) \;=\; C\,\frac{p^2 \,+\, l\,p^{\eta} \,+\, s}
                         {p^4 \,+\, u^2 \, p^2 \,+\, t^2} \,,
\label{eq:f2dgluon}
\end{equation}
which is a simple generalization of Eq.\ (\ref{eq:f4dgluon}). Results of
the fit for the various lattice volumes, using unimproved and improved
momenta, are reported in Tables \ref{tab:gluon2d-unim} and
\ref{tab:gluon2d-im} respectively.\footnote{Note that in this case the
use of improved momenta does not affect the quality of the fit significantly.
Nevertheless, we choose to consider improved momenta in our analysis also
in 2d.}
A plot of the fit for the lattice volume $V = 320^2$ using improved
momenta can be seen in Fig.\ \ref{fig:D2d}.

It is interesting to note that the function $f_{7}(p^2)$ above can be
decomposed as
\begin{equation}
f_{8}(p^2) \;=\; \frac{\alpha_+(p^2)}{p^2 + \omega_+^2} \, + \,
                 \frac{\alpha_-(p^2)}{p^2 + \omega_-^2}
           \;=\; \frac{2 a\, p^2 \,+\, 2 c w\, p^{\eta} \,+\, 2 (a v + b w)}
                      {p^4 \,+\, 2 v\, p^2 \,+\, v^2 + w^2}\,,
\label{eq:fpoles2d}
\end{equation}
with
\begin{equation}
\alpha_{\pm}(p^2) = a \pm i (b + c p^{\eta})\,, \quad\quad
\omega_{\pm}^2 = v \pm i w\,.
\label{eq:alphadef}
\end{equation}
An estimate for these five parameters is
reported in Table \ref{tab:gluon2d-param}, using again three different
analyses for the error. The average values of the fit parameters in
Eq.\ (\ref{eq:f2dgluon}) using the bootstrap method for the
lattice volume $V = 320^2$ are
\begin{eqnarray}
C & = & 0.112 \pm 0.001 \; \text{GeV}^2 \\
u & = & 0.550 \pm 0.013  \; \text{GeV}   \\
t & = & 0.255 \pm 0.006 \; \text{GeV}^2 \\
s & = & 0.0152 \pm 0.0008 \; \text{GeV}^2 \\
l & = & 0.326  \pm 0.033  \; \text{GeV}^{2-{\eta}} \\
{\eta} & = & 0.859  \pm 0.026  \,.
\label{eq:e2d}
\end{eqnarray}

\begin{table}
\begin{center}
\begin{tabular}{cccccccc}
\hline
\hline
$ V $ &  $C (\text{GeV}^2)$  &   $ u (\text{GeV})$  &  $t (\text{GeV}^2)$  &
$ s (\text{GeV}^2)$ & $l (\text{GeV}^{2-\eta})$ & $\eta$ & $\chi^2$/d.o.f.\ \\
\hline
$ 80^2$ & 0.073 (0.005) & 0.363 (0.041) & 0.265 (0.011) & 0.078 (0.008) & 0.403 (0.142) & 1.129 (0.151) & 2.92 \\
\hline
$120^2$ & 0.069 (0.004) & 0.432 (0.027) & 0.252 (0.008) & 0.052 (0.005) & 0.566 (0.128) & 1.145 (0.088) & 2.75 \\
\hline
$160^2$ & 0.067 (0.004) & 0.458 (0.022) & 0.250 (0.007) & 0.044 (0.003) & 0.665 (0.117) & 1.138 (0.066) & 2.68 \\
\hline
$200^2$ & 0.068 (0.003) & 0.470 (0.022) & 0.254 (0.007) & 0.037 (0.003) & 0.653 (0.104) & 1.088 (0.057) & 3.22 \\
\hline
$240^2$ & 0.069 (0.002) & 0.469 (0.018) & 0.252 (0.005) & 0.031 (0.002) & 0.626 (0.076) & 1.051 (0.041) & 2.61 \\
\hline
$280^2$ & 0.069 (0.002) & 0.483 (0.016) & 0.261 (0.005) & 0.029 (0.002) & 0.648 (0.064) & 0.994 (0.033) & 2.34 \\
\hline
$320^2$ & 0.070 (0.002) & 0.483 (0.016) & 0.260 (0.005) & 0.025 (0.002) & 0.630 (0.062) & 0.981 (0.032) & 2.77 \\
\hline
\hline
\end{tabular}
\end{center}
\caption{
Fits of the gluon-propagator data in the 2d case, for different lattice
volumes, using the fitting function $f_7(p^2)$ in Eq.\ (\ref{eq:f2dgluon})
and unimproved momenta [see Eq.\ (\ref{eq:kunim})].
We report, besides the value of the fit parameters, the $\chi^2$/d.o.f.\
obtained in each case. The whole range of momenta was considered for the fit.
Errors shown in parentheses correspond to one standard deviation.
}
\label{tab:gluon2d-unim}
\end{table}
\begin{table}
\begin{center}
\begin{tabular}{cccccccc}
\hline
\hline
$ V $ & $C (\text{GeV}^2)$  &   $ u (\text{GeV})$  &  $t (\text{GeV}^2)$  &
$ s (\text{GeV}^2)$ & $l (\text{GeV}^{2-\eta})$ & $\eta$ & $\chi^2$/d.o.f.\ \\
\hline
$ 80^2$ & 0.114 (0.002) & 0.433 (0.031) & 0.207 (0.012) & 0.031 (0.004) & 0.026 (0.043) & 0.684 (0.594) & 2.63 \\
\hline
$120^2$ & 0.112 (0.003) & 0.486 (0.024) & 0.197 (0.008) & 0.020 (0.002) & 0.091 (0.050) & 1.003 (0.186) & 2.51 \\
\hline
$160^2$ & 0.110 (0.002) & 0.503 (0.020) & 0.199 (0.006) & 0.018 (0.001) & 0.133 (0.044) & 1.027 (0.111) & 2.31 \\
\hline
$200^2$ & 0.110 (0.002) & 0.523 (0.020) & 0.201 (0.007) & 0.015 (0.001) & 0.157 (0.046) & 1.017 (0.092) & 2.92 \\
\hline
$240^2$ & 0.110 (0.002) & 0.519 (0.018) & 0.201 (0.006) & 0.013 (0.001) & 0.152 (0.038) & 0.955 (0.073) & 3.06 \\
\hline
$280^2$ & 0.110 (0.002) & 0.530 (0.016) & 0.208 (0.006) & 0.012 (0.001) & 0.168 (0.033) & 0.904 (0.055) & 2.77 \\
\hline
$320^2$ & 0.110 (0.001) & 0.539 (0.015) & 0.209 (0.006) & 0.011 (0.001) & 0.180 (0.033) & 0.909 (0.049) & 2.91 \\
\hline
\hline
\end{tabular}
\end{center}
\caption{
Fits of the gluon-propagator data in the 2d case, for different lattice
volumes, using the fitting function $f_7(p^2)$ in Eq.\ (\ref{eq:f2dgluon})
and improved momenta [see Eq.\ (\ref{eq:kim})].
We report, besides the value of the fit parameters, the $\chi^2$/d.o.f.\
obtained in each case. The whole range of momenta was considered for the fit.
Errors shown in parentheses correspond to one standard deviation.
}
\label{tab:gluon2d-im}
\end{table}

\begin{table}
\begin{center}
\begin{tabular}{cccc}
\hline
\hline
parameter  $\;$&$\;$ propagation of error $\;$ & $\;$ Monte Carlo analysis $\;$ & $\;$ bootstrap analysis \\
\hline
 $a \,(\text{GeV}^2)$ & 0.0550 (5) & 0.0550 (5) & 0.0559 (7) \\
\hline
 $b \,(\text{GeV}^2)$ & $-0.049 (8)$ & $-0.049 (7)$ & $-0.037 (2)$ \\
\hline
 $c \,(\text{GeV}^{2-\eta})$ & 0.07 (1) & 0.07 (1) & 0.089 (8)  \\
\hline
 $v \,(\text{GeV}^2)$ & 0.145 (8) & 0.145 (8) & 0.151 (7) \\
\hline
 $w \,(\text{GeV}^2)$ & 0.15 (2) & 0.15 (1) & 0.205 (6) \\
\hline
\hline
\end{tabular}
\end{center}
\caption{
Estimates of the parameters of the function $f_{8}(p^2)$ [see Eq.\
(\ref{eq:fpoles2d})] from fits (see last row of Table \ref{tab:gluon2d-im}
above) to the equivalent form $f_7(p^2)$ in Eq.\ (\ref{eq:f2dgluon}).
Errors are obtained using propagation of error, a Monte Carlo analysis with
10000 samples and a bootstrap analysis with 500 samples. In all cases we
considered the volume $V = 320^2$ and improved momenta.
Errors shown in parentheses correspond to one standard deviation.
Note that the value of $\eta$ can be obtained from the last row of Table
\ref{tab:gluon2d-im} for propagation of error and Monte Carlo analysis, and
from Eq.\ \ref{eq:e2d}, for the bootstrap analysis.
}
\label{tab:gluon2d-param}
\end{table}

We also tried an extrapolation to the infinite-volume limit of the gluon
propagator at zero momentum $D(0)$ using the function $A + B / N^{\nu}$,
where $N$ is the lattice side in lattice units.
This gives\footnote{Recall that, with our notation, the 2d gluon propagator
is dimensionless.}
\begin{eqnarray}
A & = & -0.002 \pm 0.010 \\
B & = & 1.7 \pm 0.8 \\
\nu & = & 0.7 \pm 0.1 \; ,
\end{eqnarray}
with $\chi^2$/d.o.f.\ = 1.07 (considering 7 data points). The fit improves
if one sets $A=0$. Indeed, in this case we find
\begin{eqnarray}
B & = & 1.9 \pm 0.2 \\
\nu & = & 0.71 \pm 0.02
\end{eqnarray}
with $\chi^2$/d.o.f.\ = 0.87. Thus, the value of the parameter $s$ in $f_7(p^2)$
[see Eq.\ (\ref{eq:f2dgluon})] is consistent with zero at infinite volume,
implying $D(0)=0$ in the same limit. One should note, however, that the
condition $D(0)=0$ is not obtained here with $b=0$ and $v=0$ [see
Eqs.\ (\ref{eq:fpoles2d}) and (\ref{eq:alphadef})], i.e.\ with purely
imaginary poles as in a Gribov-like propagator (\ref{gluonprop2}),
but it seems to be due
to the relations $a = -b$ and $v=w$ (see Table \ref{tab:gluon2d-param}).
Thus, the behavior in the 2d case appears closer to the RGZ propagator
(\ref{prop}) (with a non-integer power in the numerator)
than to a GZ propagator (\ref{gluonprop2}).
However, as mentioned in the Introduction, one
cannot relate the fitting parameters to dimension-two condensates in
the 2d case.

Finally, taking into account the infinite-volume limit, we set
$s=0$ in $f_7(p^2)$. In this case, the gluon propagator at small momenta
behaves as $D(p^2) \approx C l\,p^{\eta}/t^2$. From the conformal
relation (see for example Ref.\ \cite{Maas:2007uv} with d=2)
\begin{equation}
D(p^2) \,\sim\, (p^2)^{2 \kappa + (4-d)/2 - 1} \,=\, p^{4\kappa}
\end{equation}
we find the IR exponent $\kappa \approx 0.225$ using ${\eta} \approx 0.9$
(see Table \ref{tab:gluon2d-im}). This result is in reasonable agreement
with the numerical estimate in Ref.\ \cite{Maas:2007uv}.

%%%%%%%%%%%%%%%%%%%%%%%%%%%%%%%%%%%%%%%%%%%%%%%%%%%%%%%%%%%%%%%%%%%%%%%%%%%%

\section{Conclusions}
\label{conclusions}

We have performed a systematic fitting analysis of Landau-gauge
gluon-propagator data for SU(2) gauge theory in 2, 3, and 4
space-time dimensions. The fit results were matched to analytic
predictions from the RGZ framework, with the intent of calculating
physical values for the dimension-two condensates in the theory,
as well as to test the predictions in a neutral way. Indeed, as
mentioned in the Introduction, our strategy has been firstly to find
fits with convenient free parameters in simple (rational,
except for 2d) forms and then to interpret the parameters in terms
of physical quantities.
In that way, we do not bias the fits by imposing
relations between fit parameters from the predicted forms.
In particular, a direct fit to the general RGZ formula
[see Eq.\ (\ref{refinedgluonprop})] would involve six
parameters (corresponding to the four condensates, the parameter $\lambda^4$
and the overall normalization) with prescribed relations among them.
On the contrary, our best fits in the 4d and 3d cases involve, respectively,
four and five free parameters (including the overall normalization),
considering all generated data points.\footnote{We recall, however,
the importance of considering improved momenta [see Eq.\ (\ref{eq:kim})]
for the fits.}
We find that the resulting fits in 3d and 4d agree remarkably
well with the predictions from the RGZ scenario.

In particular, the 4d results are well described by the simplified
version of the RGZ gluon propagator in Eq.\ (\ref{prop}), equivalent to the
simplest Gribov-Stingl form. This corresponds to a pair of complex-conjugate
poles, as opposed to the Gribov propagator, in which the poles would be
purely imaginary. Our fit results --- using improved momenta and the
fitting function $f_1(p^2)$ in Eq.\ (\ref{eq:f4dgluon}) --- are given in
Table \ref{tab:gluon4d-im}. The condensates and pole masses obtained for
the largest lattice size are given respectively in Tables
\ref{tab:gluon4d-param2} and \ref{tab:gluon4d-param}.
The values for the condensates $M^2+\rho_1$, $m^2$ and $\lambda^4$
are in agreement with the ones obtained for the SU(3) case in
Ref.\ \cite{Dudal:2010tf}, where the analysis was done using a slightly
different notation, as previously explained.
The quantitative agreement between the infrared limit of SU(2) and SU(3)
theories was observed numerically before in
\cite{Sternbeck:2007ug,Cucchieri:2007zm}.

In 3d, our fits support the more general form of the RGZ propagator
in Eq.\ (\ref{refinedgluonprop}). Fit results --- using improved momenta
and the fitting function $f_4(p^2)$ in Eq.\ (\ref{eq:fDvrgzsimple}) ---
are given in Table \ref{tab:gluon3d-im}, while condensates and pole masses
from the largest lattice size are reported in Tables
\ref{tab:gluon3d-param2} and \ref{tab:gluon3d-param}.
In this case, the condensate $\rho$ is a complex quantity and there
are significant differences in the values of the other condensates
and of $\lambda^4$ compared to the 4d case. Also, in 3d one has a real
pole mass in addition to the pair of complex-conjugate poles.
It is interesting to note that the masses from the complex poles assume
similar values in 3d and 4d,
with an imaginary part that is more than twice their real part.
(We recall that a Gribov propagator would have a null real part.)
Note also that the mass and the coefficient
associated with the real pole in 3d are very small.

\vskip 3mm
In the 2d case, two particular features arise in the analysis
of the data. Firstly, as known from previous lattice studies,
the description of the infrared behavior of the gluon propagator $D(p^2)$
requires a non-integer power $\eta$ of the momentum $p$. [See the fitting
form $f_7(p^2)$ in Eq.\ (\ref{eq:f2dgluon}) and fit parameters
obtained using improved momenta in Table \ref{tab:gluon2d-im}.]
Secondly, the pole structure
that best fits the data is similar to the one observed in the 3d and 4d
cases --- i.e.\ complex-conjugate poles with nonzero real part --- at all
considered (finite) lattice volumes. In the infinite-volume limit, one
finds $D(0) = 0$, as would be the case for a Gribov propagator, with
purely imaginary poles. However, in our case, the real part of the
poles does not seem to vanish in this limit. The null value of $D(0)$ comes,
instead, from an exact cancellation of the contributions from the two
complex-conjugate poles (see values of pole masses in Table
\ref{tab:gluon2d-param}).

\vskip 3mm
Our analysis strongly suggests --- in d = 2, 3, 4 --- a pole structure
with complex-conjugate masses (with comparable real and imaginary parts)
for the infrared gluon propagator in Landau
gauge.\footnote{Let us mention that this complex-conjugate pole structure has
been shown to describe also the longitudinal and transverse gluon propagators
in Landau gauge at finite temperature
\cite{Cucchieri:2011ga,Cucchieri:2011di,FT}, at least up to twice the
critical temperature $T_c$.} As stressed at the end of Section
\ref{sec:RGZgluon}, one can
interpret this result as describing an unstable particle
\cite{Stingl:1985hx,Zwanziger:1991gz,Stingl:1994nk}. In particular, by
considering the position $\,m^2_g - i m_g \Gamma_g\,$ of the gluon pole,
one can evaluate the gluon mass $m_g$ and its width $\Gamma_g$, which are
in principle gauge-independent quantities \cite{Nielsen:1975fs,Grassi:2001bz}.
In our case, if we take as a reference the pole masses in the 4d case ---
i.e.\ the values $\omega_{\pm}^2 = v \pm i w $ with $v \approx 0.3$ GeV$^2$
and $w \approx 0.65$ GeV$^2$ from Table \ref{tab:gluon4d-param} --- we obtain
\begin{equation}
m_g \;\approx\; 550\;\; \mbox{MeV} \qquad \quad \mbox{and} \;\qquad \quad
\Gamma_g \;\approx\; 1180\;\; \mbox{MeV}\,.
\end{equation}
Note that the value for the gluon mass $m_g$ is in agreement with other
determinations \cite{Natale:2006nv,Oliveira:2010xc,Dudal:2010tf,Boucaud:2010gr}.\footnote{Let
us remark that in Ref.\ \cite{Brodsky:2008be} the authors prefer to consider,
instead of the gluon mass $m_g$, the maximum wavelength of (confined)
gluons, roughly corresponding to the
inverse gluon mass $1/m_g$.} At the same time, the very large value for
the width $\Gamma_g$ would correspond to a lifetime $\tau_g$ smaller than
10$^{-24}$ s, supporting the existence of very short-lived excitations of
the gluon field.

\vskip 3mm
In summary, we have presented fits --- inspired by a renormalizable
action --- allowing a good description of lattice data for
the Landau-gauge SU(2) gluon propagator $D(p^2)$.
The data points range from about 4 GeV down to 20--40 MeV, which are the
smallest simulated momenta to date. Our results thus provide an accurate
modeling of $D(p^2)$ in the whole IR region, which will hopefully be
a useful input in general studies of the IR sector of
Yang-Mills theories.

%%%%%%%%%%%%%%%%%%%%%%%%%%%%%%%%%%%%%%%%%%%%%%%%%%%%%%%%%%%%%%%%%%%%%%%%%%%%

\section*{Acknowledgments}

D.~Dudal and N.~Vandersickel are supported by the Research-Foundation
Flanders (FWO). A.~Cucchieri and T.~Mendes thank CNPq and FAPESP for
partial support. A.~Cucchieri also acknowledges financial support from
the Special Research Fund of Ghent University (BOF UGent).

%%%%%%%%%%%%%%%%%%%%%%%%%%%%%%%%%%%%%%%%%%%%%%%%%%%%%%%%%%%%%%%%%%%%%%%%%%%%


\begin{thebibliography}{99}

\bibitem{Greensite:2003bk}
  J.~Greensite,
  %``The Confinement problem in lattice gauge theory,''
  Prog.\ Part.\ Nucl.\ Phys.\  {\bf 51}, 1 (2003).
  %[hep-lat/0301023].

\bibitem{Gribov:1977wm}
  V.~N.~Gribov,
  %``Quantization of Nonabelian Gauge Theories,''
  Nucl.\ Phys.\  {\bf B139}, 1 (1978).

\bibitem{Zwanziger:1991gz}
  D.~Zwanziger,
  %``Vanishing of zero momentum lattice gluon propagator and color
  %confinement,''
  Nucl.\ Phys.\  {\bf B364}, 127 (1991).
  %%CITATION = NUPHA,B364,127;%%

\bibitem{Dudal:2008sp}
  D.~Dudal, J.~A.~Gracey, S.~P.~Sorella, N.~Vandersickel, H.~Verschelde,
  %``A Refinement of the Gribov-Zwanziger approach in the Landau gauge: Infrared propagators in harmony with the lattice results,''
  Phys.\ Rev.\  {\bf D78}, 065047 (2008).

\bibitem{Cucchieri:2007md}
  A.~Cucchieri, T.~Mendes,
  %``What's up with IR gluon and ghost propagators in Landau gauge? A puzzling answer from huge lattices,''
  PoS {\bf LAT2007}, 297 (2007).

\bibitem{Cucchieri:2007rg}
  A.~Cucchieri, T.~Mendes,
  %``Constraints on the IR behavior of the gluon propagator in Yang-Mills theories,''
  Phys.\ Rev.\ Lett.\  {\bf 100}, 241601 (2008).

\bibitem{Cucchieri:2008fc}
  A.~Cucchieri, T.~Mendes,
  %``Constraints on the IR behavior of the ghost propagator in Yang-Mills theories,''
  Phys.\ Rev.\  {\bf D78}, 094503 (2008).

\bibitem{Cucchieri:2010xr}
  A.~Cucchieri and T.~Mendes,
  %``Numerical test of the Gribov-Zwanziger scenario in Landau gauge,''
  PoS {\bf QCD-TNT09}, 026 (2009).
  %[arXiv:1001.2584 [hep-lat]].
  %%CITATION = POSCI,QCD-TNT09,026;%%

\bibitem{preparation}
  A.~Cucchieri, D.~Dudal, T.~Mendes and N.~Vandersickel, in preparation.

\bibitem{Aguilar:2011yb}
  A.~C.~Aguilar, D.~Binosi, J.~Papavassiliou,
  %``Gluon mass through ghost synergy,''
  [arXiv:1108.5989 [hep-ph]].

\bibitem{Zwanziger:1992qr}
  D.~Zwanziger,
  %``Renormalizability of the critical limit of lattice gauge theory by BRS invariance,''
  Nucl.\ Phys.\  {\bf B399}, 477 (1993).

\bibitem{Zwanziger:2002sh}
  D.~Zwanziger,
  %``No confinement without Coulomb confinement,''
  Phys.\ Rev.\ Lett.\  {\bf 90}, 102001 (2003).
  %[hep-lat/0209105].

\bibitem{Zwanziger:1989mf}
  D.~Zwanziger,
  %``Local And Renormalizable Action From The Gribov Horizon,''
  Nucl.\ Phys.\  {\bf B323}, 513 (1989).

\bibitem{Zwanziger:1993dh}
  D.~Zwanziger,
  %``Fundamental modular region, Boltzmann factor and area law in lattice gauge theory,''
  Nucl.\ Phys.\  {\bf B412}, 657 (1994).

\bibitem{Bogolubsky:2007ud}
  I.~L.~Bogolubsky, E.~M.~Ilgenfritz, M.~Muller-Preussker, A.~Sternbeck,
  %``The Landau gauge gluon and ghost propagators in 4D SU(3) gluodynamics in large lattice volumes,''
  PoS {\bf LAT2007}, 290 (2007).

\bibitem{Sternbeck:2007ug}
  A.~Sternbeck, L.~von Smekal, D.~B.~Leinweber, A.~G.~Williams,
  %``Comparing SU(2) to SU(3) gluodynamics on large lattices,''
  PoS {\bf LAT2007}, 340 (2007).

\bibitem{Bogolubsky:2009dc}
  I.~L.~Bogolubsky, E.~M.~Ilgenfritz, M.~Muller-Preussker, A.~Sternbeck,
  %``Lattice gluodynamics computation of Landau gauge Green's functions in the deep infrared,''
  Phys.\ Lett.\  {\bf B676}, 69 (2009).

\bibitem{Bornyakov:2009ug}
  V.~G.~Bornyakov, V.~K.~Mitrjushkin, M.~Muller-Preussker,
  %``SU(2) lattice gluon propagator: Continuum limit, finite-volume effects and infrared mass scale m(IR),''
  Phys.\ Rev.\  {\bf D81}, 054503 (2010).

\bibitem{Cucchieri:2004mf}
  A.~Cucchieri, T.~Mendes and A.~R.~Taurines,
  %``Positivity violation for the lattice Landau gluon propagator,''
  Phys.\ Rev.\  {\bf D71}, 051902 (2005).
  %[arXiv:hep-lat/0406020].

\bibitem{Fischer:2008uz}
  C.~S.~Fischer, A.~Maas, J.~M.~Pawlowski,
  %``On the infrared behavior of Landau gauge Yang-Mills theory,''
  Annals Phys.\  {\bf 324}, 2408 (2009).

\bibitem{Binosi:2009qm}
  D.~Binosi, J.~Papavassiliou,
  %``Pinch Technique: Theory and Applications,''
  Phys.\ Rept.\  {\bf 479}, 1 (2009).

\bibitem{Iritani:2009mp}
  T.~Iritani, H.~Suganuma, H.~Iida,
  %``Gluon-propagator functional form in the Landau gauge in SU(3) lattice QCD:
  %  Yukawa-type gluon propagator and anomalous gluon spectral function,''
  Phys.\ Rev.\  {\bf D80}, 114505 (2009).

\bibitem{Dudal:2010tf}
  D.~Dudal, O.~Oliveira, N.~Vandersickel,
  %``Indirect lattice evidence for the Refined Gribov-Zwanziger formalism and the gluon condensate $\braket{A^2}$ in the Landau gauge,''
  Phys.\ Rev.\  {\bf D81}, 074505 (2010).

\bibitem{Aguilar:2010gm}
  A.~C.~Aguilar, D.~Binosi, J.~Papavassiliou,
  %``QCD effective charges from lattice data,''
  JHEP {\bf 1007}, 002 (2010).

\bibitem{Cucchieri:2009zt}
  A.~Cucchieri, T.~Mendes,
  %``Landau-gauge propagators in Yang-Mills theories at beta = 0: Massive solution versus conformal scaling,''
  Phys.\ Rev.\  {\bf D81}, 016005 (2010).
%  [arXiv:0904.4033 [hep-lat]].

\bibitem{Tissier:2010ts}
  M.~Tissier, N.~Wschebor,
  %``Infrared propagators of Yang-Mills theory from perturbation theory,''
  Phys.\ Rev.\  {\bf D82}, 101701 (2010).

\bibitem{Pennington:2011xs}
  M.~R.~Pennington, D.~J.~Wilson,
  %``Are the Dressed Gluon and Ghost Propagators in the Landau Gauge presently determined in the confinement regime of QCD?,''
  [arXiv:1109.2117 [hep-ph]].

\bibitem{Oliveira:2010xc}
  O.~Oliveira, P.~Bicudo,
  %``Running Gluon Mass from Landau Gauge Lattice QCD Propagator,''
  J.\ Phys.\ {\bf G38}, 045003 (2011).

\bibitem{Aguilar:2010zx}
  A.~C.~Aguilar, D.~Binosi, J.~Papavassiliou,
  %``Nonperturbative gluon and ghost propagators for d=3 Yang-Mills,''
  Phys.\ Rev.\  {\bf D81}, 125025 (2010).

\bibitem{Tissier:2011ey}
  M.~Tissier, N.~Wschebor,
  %``An Infrared Safe perturbative approach to Yang-Mills correlators,''
  Phys.\ Rev.\  {\bf D84}, 045018 (2011).
  %[arXiv:1105.2475 [hep-th]].

\bibitem{Aiso:1997au}
  H.~Aiso {\it et al.}, % J.~Fromm, M.~Fukuda, T.~Iwamiya, A.~Nakamura, M.~Stingl, M.~Yoshida,
  %``Towards understanding of confinement of gluons,''
  Nucl.\ Phys.\ Proc.\ Suppl.\  {\bf 53}, 570 (1997).

\bibitem{Leinweber:1998uu}
  D.~B.~Leinweber, J.~I.~Skullerud, A.~G.~Williams and C.~Parrinello  [UKQCD Collaboration],
  %``Asymptotic scaling and infrared behavior of the gluon propagator,''
  Phys.\ Rev.\  {\bf D60}, 094507 (1999)
  [Erratum-ibid.\  {\bf D61}, 079901 (2000)].

\bibitem{Mandula:1999nj}
  J.~E.~Mandula,
  %``The gluon propagator,''
  Phys.\ Rept.\  {\bf 315}, 273 (1999).

\bibitem{Cucchieri:2003di}
  A.~Cucchieri, T.~Mendes, A.~R.~Taurines,
  %``SU(2) Landau gluon propagator on a 140**3 lattice,''
  Phys.\ Rev.\  {\bf D67}, 091502 (2003).

\bibitem{Stingl:1985hx}
  M.~Stingl,
  %``Propagation Properties And Condensate Formation Of The Confined
  %Yang-mills Field,''
  Phys.\ Rev.\  {\bf D34}, 3863 (1986) [Erratum-ibid. D 36, 651 (1987)].

\bibitem{Stingl:1994nk}
  M.~Stingl,
  %``A Systematic extended iterative solution for quantum chromodynamics,''
  Z.\ Phys.\  {\bf A353}, 423 (1996).
  %[hep-th/9502157].

\bibitem{Dudal:2008rm}
  D.~Dudal, J.~A.~Gracey, S.~P.~Sorella, N.~Vandersickel, H.~Verschelde,
  %``The Landau gauge gluon and ghost propagator in the refined Gribov-Zwanziger framework in 3 dimensions,''
  Phys.\ Rev.\  {\bf D78}, 125012 (2008).

\bibitem{Vandersickel:2011ye}
  N.~Vandersickel, D.~Dudal, S.~P.~Sorella,
  %``More evidence for a refined Gribov-Zwanziger action based on an effective potential approach,''
  PoS {\bf FACESQCD}, 044 (2010).
  %[arXiv:1102.0866 [hep-th]].

\bibitem{Vandersickel:2011zc}
  N.~Vandersickel,
  %``A Study of the Gribov-Zwanziger action: from propagators to glueballs,''
  [arXiv:1104.1315 [hep-th]].

\bibitem{Dudal:2011gd}
  D.~Dudal, S.~P.~Sorella, N.~Vandersickel,
  %``The dynamical origin of the refinement of the Gribov-Zwanziger theory,''
  Phys.\ Rev.\  {\bf D84}, 065039 (2011).
  %[arXiv:1105.3371 [hep-th]].

\bibitem{Dudal:2008xd}
  D.~Dudal, S.~P.~Sorella, N.~Vandersickel, H.~Verschelde,
  %``The Effects of Gribov copies in 2D gauge theories,''
  Phys.\ Lett.\  {\bf B680}, 377 (2009).
  %[arXiv:0808.3379 [hep-th]].

\bibitem{Maggiore:1993wq}
  N.~Maggiore, M.~Schaden,
  %``Landau gauge within the Gribov horizon,''
  Phys.\ Rev.\  {\bf D50 }, 6616 (1994).
  % [hep-th/9310111].

\bibitem{Dudal:2010fq}
  D.~Dudal, S.~P.~Sorella, N.~Vandersickel,
  %``More on the renormalization of the horizon function of the Gribov-Zwanziger action and the Kugo-Ojima Green function(s),''
  Eur.\ Phys.\ J.\  {\bf C68}, 283 (2010).

\bibitem{Sobreiro:2004us}
  R.~F.~Sobreiro, S.~P.~Sorella, D.~Dudal, H.~Verschelde,
  %``Gribov horizon in the presence of dynamical mass generation in Euclidean Yang-Mills theories in the Landau gauge,''
  Phys.\ Lett.\  {\bf B590}, 265 (2004).

\bibitem{Zwanziger:1990by}
  D.~Zwanziger,
  %``Vanishing color magnetization in lattice Landau and Coulomb gauges,''
  Phys.\ Lett.\  {\bf B257}, 168 (1991).

\bibitem{Gracey:2007vv}
  J.~A.~Gracey,
  %``Exploring the infrared structure of QCD with the Gribov-Zwanziger Lagrangian,''
  Braz.\ J.\ Phys.\  {\bf 37}, 226 (2007).

\bibitem{Ford:2009ar}
  F.~R.~Ford, J.~A.~Gracey,
  %``Two loop anti-MS Gribov mass gap equation with massive quarks,''
  J.\ Phys.\ {\bf A42}, 325402 (2009).

\bibitem{Zwanziger:2010iz}
  D.~Zwanziger,
  %``Goldstone bosons and fermions in QCD,''
  Phys.\ Rev.\  {\bf D81}, 125027 (2010).
  %[arXiv:1003.1080 [hep-ph]].

\bibitem{Dokshitzer:2004ie}
  Y.~L.~Dokshitzer, D.~E.~Kharzeev,
  %``The Gribov conception of quantum chromodynamics,''
  Ann.\ Rev.\ Nucl.\ Part.\ Sci.\  {\bf 54}, 487 (2004).
  %[hep-ph/0404216].

\bibitem{Alkofer:2003jj}
  R.~Alkofer, W.~Detmold, C.~S.~Fischer, P.~Maris,
  %``Analytic properties of the Landau gauge gluon and quark propagators,''
  Phys.\ Rev.\  {\bf D70 } (2004)  014014.
  %[hep-ph/0309077].

\bibitem{Bernard:1992hy}
  C.~W.~Bernard, C.~Parrinello, A.~Soni,
  %``The Gluon propagator in momentum space,''
  Nucl.\ Phys.\ Proc.\ Suppl.\  {\bf 30}, 535 (1993).
  %[hep-lat/9211020].

\bibitem{Cucchieri:2006za}
  A.~Cucchieri, T.~Mendes,
  %``Infrared behavior of gluon and ghost propagators from asymmetric lattices,''
  Phys.\ Rev.\  {\bf D73}, 071502 (2006).
  %[arXiv:hep-lat/0602012 [hep-lat]].

\bibitem{lcca}
  {\tt http://www.usp.br/lcca/IBM.html}.

\bibitem{Cucchieri:1995pn}
  A.~Cucchieri, T.~Mendes,
  %``Critical slowing down in SU(2) Landau gauge fixing algorithms,''
  Nucl.\ Phys.\  {\bf B471}, 263 (1996).

\bibitem{Bloch:2003sk}
  J.~C.~R.~Bloch, A.~Cucchieri, K.~Langfeld, T.~Mendes,
  %``Propagators and running coupling from SU(2) lattice gauge theory,''
  Nucl.\ Phys.\  {\bf B687}, 76 (2004).

\bibitem{Maas:2007uv}
  A.~Maas,
  %``Two and three-point Green's functions in two-dimensional Landau-gauge Yang-Mills theory,''
  Phys.\ Rev.\  {\bf D75}, 116004 (2007).

\bibitem{Cucchieri:1997dx}
  A.~Cucchieri,
  %``Gribov copies in the minimal Landau gauge: The influence on gluon and
  %ghost propagators,''
  Nucl.\ Phys.\  {\bf B508}, 353 (1997).

\bibitem{Silva:2004bv}
  P.~J.~Silva and O.~Oliveira,
  %``Gribov copies, lattice QCD and the gluon propagator,''
  Nucl.\ Phys.\  {\bf B690}, 177 (2004).

\bibitem{Bogolubsky:2005wf}
  I.~L.~Bogolubsky, G.~Burgio, M.~Muller-Preussker and V.~K.~Mitrjushkin,
  %``Landau gauge ghost and gluon propagators in SU(2) lattice gauge theory:
  %Gribov ambiguity revisited,''
  Phys.\ Rev.\  {\bf D74}, 034503 (2006).

\bibitem{Bogolubsky:2009qb}
  I.~L.~Bogolubsky, E.~M.~Ilgenfritz, M.~Muller-Preussker and A.~Sternbeck,
  %``The Landau gauge gluon propagator in 4D SU(2) lattice gauge theory
  %revisited: Gribov copies and scaling properties,''
  PoS {\bf LATTICE2009}, 237 (2009).

\bibitem{Maas:2009ph}
  A.~Maas, J.~M.~Pawlowski, D.~Spielmann, A.~Sternbeck, L.~von Smekal,
  %``Strong-coupling study of the Gribov ambiguity in lattice Landau gauge,''
  Eur.\ Phys.\ J.\  {\bf C68}, 183-195 (2010).

\bibitem{Furui:2005bu}
  S.~Furui and H.~Nakajima,
  %``Infrared features of KS fermion and Wilson fermion in Lattice Landau Gauge
  %QCD,''
  Few Body Syst.\  {\bf 40}, 101 (2006).

\bibitem{Ilgenfritz:2006he}
  E.~M.~Ilgenfritz, M.~Muller-Preussker, A.~Sternbeck, A.~Schiller and I.~L.~Bogolubsky,
  %``Landau gauge gluon and ghost propagators from lattice QCD,''
  Braz.\ J.\ Phys.\  {\bf 37}, 193 (2007).

\bibitem{Bowman:2007du}
  P.~O.~Bowman {\it et al.},
  % P.~O.~Bowman, U.~M.~Heller, D.~B.~Leinweber, M.~B.~Parappilly, A.~Sternbeck, L.~von Smekal, A.~G.~Williams, J.~-b.~Zhang,
  %``Scaling behavior and positivity violation of the gluon propagator in full
  %QCD,''
  Phys.\ Rev.\  {\bf D76}, 094505 (2007).

\bibitem{Boucaud:2001un}
  P.~Boucaud {\it et al.},
  % J.~P.~Leroy, J.~Micheli, H.~Moutarde, O.~Pene, J.~Rodriguez-Quintero and C.~Roiesnel,
  %``Unquenched calculation of alpha(s) from Green functions: Progress
  %report,''
  Nucl.\ Phys.\ Proc.\ Suppl.\  {\bf 106}, 266 (2002).

\bibitem{Cucchieri:2003zx}
  A.~Cucchieri, T.~Mendes, G.~Travieso, A.~R.~Taurines,
  %``Parallel implementation of a lattice gauge theory code: Studying quark confinement on PC clusters,''
  15th Symposium on Computer Architecture and High Performance Computing (SBAC-PAD'03), 123 (2003)
  [hep-lat/0308005].

\bibitem{Cucchieri:1999sz}
  A.~Cucchieri,
  %``Infrared behavior of the gluon propagator in lattice Landau gauge: The Three-dimensional case,''
  Phys.\ Rev.\  {\bf D60}, 034508 (1999).

\bibitem{Ma:1999kn}
  J.~P.~Ma,
  %``A study of gluon propagator on coarse lattice,''
  Mod.\ Phys.\ Lett.\  {\bf A15}, 229 (2000).

\bibitem{deSoto:2007ht}
  F.~de Soto and C.~Roiesnel,
  %``On the reduction of hypercubic lattice artifacts,''
  JHEP {\bf 0709}, 007 (2007).

\bibitem{gnuplot} {\tt http://www.gnuplot.info/documentation.html}.

\bibitem{Cucchieri:2011um}
  A.~Cucchieri, T.~Mendes,
  %``The Saga of Landau-Gauge Propagators: Gathering New Ammo,''
  AIP Conf.\ Proc.\  {\bf 1343}, 185 (2011).
  %[arXiv:1101.4779 [hep-lat]].

\bibitem{Cucchieri:2011ga}
  A.~Cucchieri, T.~Mendes,
  %``Further Investigation of Massive Landau-Gauge Propagators in the Infrared Limit,''
  PoS {\bf LATTICE2010}, 280 (2010).

\bibitem{Cucchieri:2011di}
  A.~Cucchieri, T.~Mendes,
  %``Electric and magnetic Landau-gauge gluon propagators in finite-temperature SU(2) gauge theory,''
  PoS {\bf FACESQCD}, 007 (2010).
  %[arXiv:1105.0176 [hep-lat]].

\bibitem{FT}
  A.~Cucchieri, T.~Mendes, in preparation.

\bibitem{Cucchieri:2007zm}
  A.~Cucchieri, T.~Mendes, O.~Oliveira, P.~J.~Silva,
  %``Just how different are SU(2) and SU(3) Landau propagators in the IR regime?,''
  Phys.\ Rev.\  {\bf D76}, 114507 (2007).
  %[arXiv:0705.3367 [hep-lat]].

\bibitem{Nielsen:1975fs}
  N.~K.~Nielsen,
  %``On the Gauge Dependence of Spontaneous Symmetry Breaking in Gauge Theories,''
  Nucl.\ Phys.\  {\bf B101}, 173 (1975).

\bibitem{Grassi:2001bz}
  P.~A.~Grassi, B.~A.~Kniehl, A.~Sirlin,
  %``Width and partial widths of unstable particles in the light of the Nielsen identities,''
  Phys.\ Rev.\  {\bf D65}, 085001 (2002).
  %[hep-ph/0109228].

\bibitem{Natale:2006nv}
  A.~A.~Natale,
  %``Phenomenology of infrared finite gluon propagator and coupling constant,''
  Braz.\ J.\ Phys.\  {\bf 37}, 306 (2007).
  %[arXiv:hep-ph/0610256 [hep-ph]].

\bibitem{Boucaud:2010gr}
  P.~.Boucaud {\it et al.}, % M.~E.~Gomez, J.~P.~Leroy, A.~Le Yaouanc, J.~Micheli, O.~Pene, J.~Rodriguez-Quintero,
  %``The low-momentum ghost dressing function and the gluon mass,''
  Phys.\ Rev.\  {\bf D82}, 054007 (2010).
  %[arXiv:1004.4135 [hep-ph]].

\bibitem{Brodsky:2008be}
  S.~J.~Brodsky, R.~Shrock,
  %``Maximum Wavelength of Confined Quarks and Gluons and Properties of Quantum Chromodynamics,''
  Phys.\ Lett.\  {\bf B666 } (2008)  95.
  %[arXiv:0806.1535 [hep-th]].

\end{thebibliography}
\end{document}